\documentclass{iopart}

\usepackage{epsfig}
\usepackage{iopams}
\usepackage{hyperref}
\usepackage{amssymb}
\usepackage{bm}
\usepackage{bbm}
\usepackage{color}
\usepackage{multirow}
\usepackage{threeparttable}

\def\sgn{\mathop{\textrm{sgn}}}

\newcommand{\beq}{\begin{equation}}
\newcommand{\eeq}{\end{equation}}
\newcommand{\beqarray}{\begin{eqnarray}}
\newcommand{\eeqarray}{\end{eqnarray}}
  
\newcommand{\Sec}[1]{Sec.~\ref{#1}}  
\newcommand{\eq}[1]{Eq.~(\ref{#1})}  
\newcommand{\fig}[1]{Fig.~\ref{#1}}  
  
\newcommand{\eqref}[1]{(\ref{#1})}

\begin{document}

\article[Topological surface states in nodal superconductors]{}{Topological surface states in nodal superconductors}

\date{\today}

\author{Andreas P. Schnyder}
\ead{a.schnyder@fkf.mpg.de}
\address{Max-Planck-Institut f\"ur Festk\"orperforschung,
  Hei\ss{}enbergstrasse 1, D-70569 Stuttgart, Germany} 

\author{Philip M. R. Brydon}
\ead{pbrydon@umd.edu}
\address{Condensed Matter Theory Center, University of Maryland, College Park, MD 20742, USA}

\begin{abstract}
Topological superconductors have become a subject of intense research
due to their potential use for technical applications in device
fabrication and quantum information.  
Besides fully gapped superconductors, unconventional superconductors
with point or line nodes in their order parameter can also exhibit
nontrivial topological characteristics. 
This article reviews recent progress in the theoretical  understanding
of nodal topological superconductors, with 
a focus on Weyl and noncentrosymmetric superconductors and
their protected surface states. Using selected examples, we review the
bulk topological properties of these systems,  
study  different types of topological surface states, and examine
their unusual properties. 
Furthermore, we survey some candidate materials for topological
superconductivity and 
discuss different experimental signatures of topological surface states.
\end{abstract}

\date{\today}

\pacs{03.65.vf, 74.45.+c,74.20.Rp,74.25.F-,73.20.-r}

% 03.65.vf Topological phase / Berry phase
% 74.45.+c Superconductivity Andreev reflection
% 74.50.+r Josephson effect, tunelling phenomena
% 74.20.Rp Superconductivity, pairing symmetry
% 74.25.F- Transport properties in superconductors
% 73.43.-f  Quantum Hall effect
% 72.25.Dc spin polarized transport in
% 73.20.-r Electron DOS - surfaces and interfaces, 
% 73.20.Hb Impurities - at surfaces and interfaces
% 75.70.Rf Magnetic properties -  surfaces (surface magnetism)
% 68.35.Rh Order disorder transition - at surfaces and interfaces
% 68.35.Rh Phase transition - at surfaces and interfaces, 
% 73.20.-r Surface states, 

\maketitle

\tableofcontents

\section{Introduction}

\noindent
The discovery that gapped electronic systems can possess nontrivial
topology has sparked a revolution in condensed matter physics~\cite{schnyderPRB08,kitaev22,schnyderAIP,ryuNJP2010,hasanKaneRMP,qiZhangRMP11,grinevichVolovikJLow88,saloomaVolovikPRB88}. Such
topological systems are characterized by an $\mathbb{Z}$ or
$\mathbb{Z}_2$ topological number,  and display topologically protected
surface states due to the so-called bulk-boundary correspondence. Two
species of such systems can be discerned: topological insulators and
superconductors. While there are now many examples of topological
insulators, topological superconductors have proved to be much
rarer. This is due to the unconventional pairing symmetries
required for a topological state. In contrast to the familiar $s$-wave
spin-singlet pairing in conventional superconductors, such pairing
states appear to require a particular
constellation of pairing interactions and electronic structure, and
are not robust to disorder. Topological superconductors are
nevertheless of great interest due to the existence of protected
Majorana surface states, which arise as a consequence of a nontrivial
topology of the bulk quasiparticle wave functions. These surface
states may have important applications in quantum information
technology, and have motivated an 
intense effort to ``engineer'' a topological superconducting
state from more prosaic
components~\cite{aliceaRepProg,beenakkerReview,nayakRMP08,stanescu2013}. While
there are many such 
proposals, at present the most 
promising is to proximity-induce $p$-wave superconductivity in a
semiconductor with strong spin-orbit coupling~\cite{FuKane_SC_STI,Sau_semiconductor_heterostructures,Roman_SC_semi,Gil_Majorana_wire,Mourik25052012}. 

Although superconductors with unconventional pairing symmetries are not
uncommon, they rather typically display points or lines on the Fermi surface
where the superconducting gap vanishes, so-called ``nodes''. As such,
they are not covered by the ``ten fold way'' classification of the
topology of \emph{fully gapped} electronic systems~\cite{schnyderPRB08,kitaev22,schnyderAIP,ryuNJP2010}. Nevertheless, such
systems can display interesting topological properties, and possess
topologically-protected edge states, although the protection is weaker
than for fully-gapped systems~\cite{horavaPRL05,beriPRB10,satoPRB06,Schnyder10,Schnyder12,Brydon11,yadaPRB2011,Volovik:book,VolovikExotic,volovikLectNotes13,Volovik3HeA,heikkKopninVolovik11,heikkilaVolovikJETP11,dahlhaus2012}.
 Although global topological invariants cannot be defined for gapless
 systems, it is nevertheless possible to classify  
 the rich topological structure of nodal superconductors in terms of
 momentum-dependent topological numbers.  
This has revealed that many nodal topological superconductors and
superfluids have nontrivial topology, e.g. the
high-T$_c$ $d_{x^2-y^2}$-wave superconductors
\cite{RyuHatsugaiPRL02,huPRL94,Scalapino1995329,tsueiRMP00}, the heavy Fermion systems CeCoIn$_5$
\cite{IzawaPRL01,MovshovichPRL01,allanDavis13,zhouYazdani13} and URu$_2$Si$_2$ \cite{kasahara07,YanoMachidaPRL08,liBalicas13,goswamiArXiv13}, 
 nodal noncentrosymmetric superconductors \cite{satoPRB06,Schnyder10,Schnyder12,Brydon11,yadaPRB2011,tanakaPRL2010,satoPRB2011}, 
 Weyl superconductors \cite{biswasPRB2013,fischerPRB14,meng12,das13},
 oxide interfaces \cite{OhtomoNature04}, and the A phase of superfluid
 ${}^3$He
 \cite{Volovik:book,VolovikExotic,volovikLectNotes13,Volovik3HeA,vollhardtBook}.  Similar to fully gapped topological superconductors, the topological characteristics of nodal superconducting states reveal themselves at the surface
 in the form of Majorana cone states, Majorana flat-band states, or
 arc surface states \cite{Schnyder12}.

Our aim in this review is to provide a thorough introduction into the
field of nodal topological superconductors, acting as a manual and
reference for both theorists and experimentalists. We do not wish to
deliver an exhaustive overview of all nodal superconductors, however,
but rather to show via selected examples the different topological
structures which can be realized and what type of unconventional
properties theses materials exhibit. While the concepts that we
discuss can be generally applied to all nodal topological systems, our
focus shall be upon systems where the superconductivity is an
intrinsic property, as opposed to engineered nodal superconducting
states in heterostructures.

We give a brief outline of the review here. We commence
in~\Sec{secModels} by introducing our two model systems: Weyl
superconductors and noncentrosymmetric superconductors, which are
prototypes for topological superconductors with point and line nodes,
respectively. In~\Sec{topoSection} we set out the topological
classification of the nodal states in terms of a ``periodic table'' of
the nodal structure, and apply this to our canonical examples. The
topologically-protected surface states are studied
in~\Sec{sec:quasiclassical} using a quasiclassical method, which forms
a useful basis for understanding the experimental signatures of the
topological surface states in~\Sec{sec:signatures}. We survey
candidate nodal topological superconductors
in~\Sec{sec:materialsurvey}, before concluding with a brief outlook for the
field.

\section{Phenomenological models}
\label{secModels}

While there are a plethora of nodal superconductors with
  nontrivial topology, it is convenient to illustrate our discussion
  with straightforward but physically relevant phenomenological
  models. Here we introduce three such systems: the $p$-wave Weyl
  superconductor, the $d$-wave Weyl superconductor, and
  noncentrosymmetric superconductors.
Our choice is motivated to exemplify line and point nodal structures,
and we shall return to these model systems throughout the review.

\subsection{Topological superconductors with point nodes}~\label{sec:topSCwpointnodes}

We consider two different phenomenological models of
superconductors/superfluids with point nodes.  The first one is the
Anderson-Brinkman-Morel (ABM)
state~\cite{andersonMorel,wheatleyRMP,vollhardtBook}, which is
believed to occur in the A phase of superfluid ${}^3$He as well as in
cold atomic and polar molecular gases~\cite{moellerGurariePRB11}.  
The ABM state is a three-dimensional superfluid of spinless (or fully spin-polarized) fermions with $(p_x \pm i p_y)$-wave pairing and is described
by the Bogoliubov-de Gennes (BdG) Hamiltonian $\mathcal{H} =
\frac{1}{2} \sum_{\bf k} \phi^{\dag}_{\bf k} \mathcal{H} ( {\bf k} )
\phi^{\phantom{\dag}}_{\bf k} $ with 
\begin{eqnarray} \label{hamAphase}
\mathcal{H} ( {\bf k} )
=
h({\bf k}) \tau_z + \frac{ \Delta_0 }{k_F} {\bf k} \cdot \left( \hat{\bf e}_1 \tau_x +   \hat{\bf e}_2 \tau_y \right) ,
\end{eqnarray}
where
$\phi_{\bf k} = ( c^{\ }_{\bf k} , c^{\dag}_{\bf k} )^{\mathrm{T}}$
and $c^{\dag}_{\bf k}$ ( $c^{\phantom{\dag}}_{\bf k}$ ) is the creation (annihilation) operator of spinless fermions with momentum ${\bf k}$.
Here, $ \tau_i$ denote the three Pauli matrices which act in
particle-hole space, $\hat{\bf l} = \hat{\bf e}_1 \times \hat{\bf
  e}_2$ represents the direction of the orbital momentum of the Cooper pair,
 $h({\bf k})$ is the normal-state dispersion, $k_F$ is the Fermi wave vector,
and $\Delta_0$ denotes the amplitude of the superconducting order parameter. 
In the following we assume a spherical Fermi surface centered at the $\Gamma$ point and a momentum independent vector $\hat{\bf l}$ pointing
along the $k_z$ axis in the Brillouin zone. Thus, 
the eigenvalues of 
Hamiltonian~\eqref{hamAphase} are $\pm \lambda ({\bf k} ) = \pm \sqrt{
  [ h ({\bf k}) ]^2 + ( \Delta_0 / k_F )^2 ( k_x^2 + k_y^2) }$, 
and the spectrum $\lambda ( {\bf k} )$ exhibits two Weyl nodes at the north and  south poles of the Fermi sphere ${\bf K}_{\pm} = ( 0,  0, \pm k_F)$.
The low-energy physics in the vicinity of the point nodes ${\bf
  K}_{\pm}$ can be captured by the anisotropic Weyl Hamiltonian 
\begin{eqnarray} \label{WeylHam}
\mathcal{H}_{\pm} ( {\bf p} ) 
=
(\Delta_0 / k_F ) ( p_x \tau_x + p_y \tau_y ) \pm \hbar v_F  p_z  \tau_z ,
\end{eqnarray}
where ${\bf p} = {\bf k} - {\bf K}_{\pm}$ and $v_F = \hbar k_F / m$
denotes the Fermi velocity.  
Hence, Eq.~\eqref{hamAphase} defines the simplest version of a Weyl superconductor with two point nodes. 
The dispersion close to the Weyl points is linear in all three directions,
leading to a density of states that increases quadratically with energy.
As opposed to Weyl points in semimetals, the particle-hole symmetry
dictates that the point nodes ${\bf K}_{\pm}$ remain pinned at zero
energy even as the chemical potential is varied. 
Each Weyl node can be characterized in terms of a chirality index,
which measures the relative handedness of the three momenta {\bf p}
with respect to the Pauli matrices in Eq.~\eqref{WeylHam}. 
For Hamiltonian $\mathcal{H}_\pm ( {\bf p} )$
the chirality of the Weyl point is $\pm 1$.
We note that while arbitrary (small) perturbations of
Hamiltonian~\eqref{WeylHam} in general lead to a shift of the point
nodes, they cannot open up a gap in the spectrum. 
We will see in Sec.~\ref{SecExampChiralP} that
the stability of the Weyl nodes is protected by a  Chern number (or
Skyrmion number), which takes on the values $\pm 1$.

As a second example of a superconductor with point nodes, we consider the three-dimensional 
spin-singlet chiral $(d_{x^2-y^2} \pm i d_{xy})$-wave state, which is likely realized in the heavy fermion system URu$_2$Si$_2$~\cite{kasahara07,YanoMachidaPRL08,liBalicas13,goswamiArXiv13} and the pnictide material SrPtAs~\cite{biswasPRB2013,fischerPRB14}. 
The BdG Hamiltonian of this chiral superconducting phase is
given by
$\mathcal{H}=\frac{1}{2}\sum_{\bf k}\Phi^\dag_{\bf k} \mathcal{H} ( {\bf k} ) \Phi^{\phantom{\dag}}_{\bf k}$, with 
\begin{eqnarray} \label{hamChiralD}
\mathcal{H} ( {\bf k} )
=
h ( {\bf k} ) \tau_z
+ \frac{\Delta_0}{k_F^2} 
\left[
(k_x^2 - k_y^2 ) \tau_x + 2 k_x k_y \tau_y
\right]
\end{eqnarray}
and the Nambu spinor $\Phi_{\bf k} = ( c^{\phantom{\dag}}_{{\bf k} \uparrow} , c^{\dag}_{-{\bf k} \downarrow}  )^{\mathrm{T}}$, where
 $c^{\dag}_{{\bf k} s}$ ($c^{\phantom{\dag}}_{{\bf k} s}$)
represents the electron creation (annihilation) operator with momentum ${\bf k}$ and spin $s$.
Observe that for this pairing symmetry the orbital angular momentum
projected along the $k_z$ axis is $\pm 2$, whereas for the $(p_x \pm i
p_y)$-wave state discussed above it is $ \pm 1$.
Similar to Eq.~\eqref{hamAphase}, we find that for a spherical Fermi surface centered at the $\Gamma$ point, Hamiltonian~\eqref{hamChiralD} exhibits 
two point nodes located at ${\bf K}_{\pm} = (0,0, \pm k_F)$, which
are double Weyl points pinned at zero energy by the particle-hole
symmetry. 
Expanding Eq.~\eqref{hamChiralD} around ${\bf K}_{\pm}$ to second
order in the momentum, we obtain
the following low-energy description of the double Weyl nodes
\begin{eqnarray} \label{doubleWeylPoints}
\mathcal{H}_{\pm}  ( {\bf p} ) 
=
 \frac{\Delta_0}{k_F^2} 
\left[
(p_x^2 - p_y^2 ) \tau_x + 2 p_x p_y \tau_y
\right]
+ \left[ 
\frac{\hbar^2}{2 m}  \left| {\bf p} \right|^2 \pm \hbar v_F p_z
 \right] \tau_z ,
\end{eqnarray}
where ${\bf p}=   {\bf k}Ê- {\bf K}_{\pm} $.
Hence, the low-energy nodal quasiparticles near ${\bf K}_{\pm}$
exhibit  linear and quadratic dispersions along the $p_z$ direction 
 and in the $p_x p_y$ plane, respectively. All in all, this anisotropic dispersion leads to a density of states which
 increases linearly with energy.
 As explained in Sec.~\ref{SecExampChiralD}, the double Weyl points \eqref{doubleWeylPoints} are protected against gap opening by the conservation of a Chern or Skyrmion number
that takes on the values $\pm 2$.

\subsection{Topological superconductors with line
  nodes}~\label{sec:topSCwlinenodes}

The prototype of a topological superconductor with line nodes are 
time-reversal-symmetric three-dimensional noncentrosymmetric
superconductors (NCS). In these 
systems, the crystal structure breaks inversion symmetry leading to
strong electric field gradients within the unit cell, which in turn
generates strong spin-orbit coupling (SOC). Furthermore, the absence of
inversion symmetry means that parity is no longer a good quantum
number, and so a mixed pairing state with both singlet and triplet
gaps is possible. 

The minimal description for such a system has the BdG
Hamiltonian $\mathcal{H} = \frac{1}{2} \sum_{\bf{k}}
\psi^{\dag}_{\bf{k}} \mathcal{H} (\bf{k}) \psi^{\ }_{\bf{k}}$, 
with
\beq \label{NCSham}
\mathcal{H} ( \bf{k} )
=
\left(\begin{array}{cc}
h ( \bf{k} ) & \Delta ( \bf{k} ) \\
\Delta^{\dag} ( \bf{k} ) & - h^{T} ( - \bf{k} )
\end{array}\right)
\eeq
and $\psi_{\bf{k} } = ( c^{\ }_{{\bf k} \uparrow}, c^{\ }_{\bf{k} \downarrow},
c^{\dag}_{- \bf{k} \uparrow}, c^{\dag}_{- \bf{k} \downarrow} )^{\mathrm{T}}$, 
where $c^{\dag }_{\bf{k}\sigma}$ ($c^{\  }_{\bf{k}\sigma}$) denotes the
electron creation 
(annihilation) operator with momentum ${\bf k}$ and spin $\sigma$.
The normal-state dispersion of the electrons in the spin basis is given by 
\beq
h ( {{\bf k}} )
=
\varepsilon_{{\bf k}} \sigma_0 + {\bf g}_{{\bf k}} \cdot \bm{\sigma }, \label{eq:HnormalNCS}
\eeq
where $\varepsilon_{{\bf k} }$ is the nonmagnetic dispersion while
${\bf g}_{{\bf k}}$  denotes the SOC potential.
${\bm{\sigma}} = ( \sigma_x, \sigma_y, \sigma_z )$ denotes the vector
of the Pauli matrices, while 
$\sigma_0$ stands for the $2 \times 2$ unit matrix. By
time-reversal-symmetry, we require that $\varepsilon_{{\bf k} }$
and ${\bf g}_{{\bf k}}$ are symmetric and antisymmetric in ${\bf k}$,
respectively. The normal-state Hamiltonian~\eq{eq:HnormalNCS} is
diagonalized in the so-called helicity basis, taking the form  
$\tilde{h} ( {\bf k}) =  \mathrm{diag} ( \xi^+_{\bf k} , \xi^-_{\bf k} )$, where
$\xi^{\pm}_{\bf k} = \varepsilon_{\bf k} \pm \left| {\bf g}_{\bf k} \right|$ is
the dispersion of the positive-helicity and negative-helicity bands,
respectively.  

The superconducting gap contains both even-parity spin-singlet and
odd-parity spin-triplet pairing potentials
\beq
\Delta ( {{\bf k}} ) =
 \left(  \psi_{\bf k} \sigma_0 + {\bf d}_{{\bf k} } \cdot \bm{\sigma} \right)
\left( i \sigma_y \right) ,
\eeq
where $\psi_{\bf k}$ and ${\bf d}_{\bf k}$ represent the spin-singlet and
spin-triplet components, respectively. In a time-reversal-symmetric
system they have the same global phase, and without loss of generality
we assume them to be real. It is well known that in the absence of
interband pairing, the superconducting transition temperature is maximized when 
the spin-triplet pairing vector ${\bf d}_{\bf k}$ is aligned with the
polarization vector ${\bf g}_{\bf k}$  
of the SOC~\cite{Frigeri04}. This implies that there is only pairing
between states on the same helicity Fermi surface, with positive and
negative helicity gaps $\Delta^{\pm}_{\bf   k} = \psi_{\bf k} \pm |{\bf
  d}_{\bf k}|$. Assuming an $s$-wave like order parameter for the
singlet pairing, we observe that while the positive-helicity Fermi
surface is  fully gapped, the negative-helicity gap may possess
nodes. We parametrize the singlet and triplet components of the
superconducting gap function as 
\begin{eqnarray}
\psi_{\bf k}  
&=&	
\Delta_s 
=
r\,  \Delta_0 ,	
\\
{\bf d}_{\bf k} 
&=&
\Delta_t {\bf l}_{\bf k} 
=
(1-r)\,
\Delta_0 
{\bf l}_{\bf k} ,
\end{eqnarray}
where ${\bf l}_{\bf k} = {\bf g}_{\bf k} / \lambda$, with $\lambda$ the
SOC strength.  
Here, $r$ parametrizes the mixed pairing state, tuning the system
between purely triplet ($r=0$) and purely singlet ($r=1$)
gaps.

\section{Topological classification of nodal superconductors}
\label{topoSection}

The topological properties of nodal superconductors (and semimetals) can be classified in a similar manner as those of 
fully gapped superconductors (and insulators)~\cite{schnyderPRB08,kitaev22,schnyderAIP}.
A nodal point or nodal line in a superconductor is topologically stable if there does not exist
any symmetry preserving mass term that opens up a full gap in the spectrum. The topological stability of these point or line
nodes is guaranteed by the conservation of a topological charge (i.e., a topological invariant), which is defined
in terms of an integral along a contour that encloses (encircles) the point node (line node).
These topological invariants can be either Chern or winding numbers, referred to as ``$\mathbb{Z}$ invariants"  (or ``$2\mathbb{Z}$ invariants"), or 
binary invariants, referred to as ``$\mathbb{Z}_2$ numbers".
In this section we briefly review the ten-fold classification of
nodal topological superconductors (and semimetals)~\cite{matsuura2012,ZhaoPRL13,ZhaoWangPRB14,shiozakiPRB14,chiuSchnyder2014}. 
This scheme classifies point and line nodes in nodal superconductors (Fermi points and Fermi lines in semimetals)
in terms of global symmetries, i.e., nonspatial symmetries that act locally in position space, 
namely, time-reversal symmetry (TRS), particle-hole symmetry (PHS), and chiral or sublattice symmetry (SLS).

In momentum space,  TRS and PHS act on the BdG (or Bloch) Hamiltonian $\mathcal{H} ( {\bf k} )$ as
\begin{equation} \label{TRSopPHSop}
T^{-1}  \mathcal{H} ( -{\bf k}) T 
= + \mathcal{H} ( {\bf k} )
\quad \textrm{and} \quad
C^{-1} \mathcal{H} (-{\bf k} ) C 
=  - \mathcal{H} ( {\bf k} ),
\end{equation}
respectively, 
where $T = \mathcal{K} U_T$ and $C = \mathcal{K} U_C $ are antiunitary operators and $\mathcal{K}$ denotes the complex conjugation operator.
Chiral symmetry, however, is implemented in terms of an anticommutation relation 
\begin{eqnarray} \label{chiralOPP}
S^{-1} \mathcal{H} ({\bf k} ) S 
&=& 
- \mathcal{H} ( {\bf k} ),
\end{eqnarray}
where $S$ is a unitary operator. Chiral symmetry $S$ can be viewed as a combination of time-reversal and particle-hole symmetry, i.e., $S \propto T C$.
There are three different possibilities for how a given BdG (or Bloch) Hamiltonian $\mathcal{H}({\bf k})$ can transform under TRS (or PHS): (i) 
$\mathcal{H} ({\bf k})$ is not symmetric under TRS (or PHS), which is denoted by ``0" in Table~\ref{classificationTable}; 
(ii) $\mathcal{H} ( {\bf k} )$ is invariant under TRS (or PHS) and $T$ (or $C$) squares to $+1$, which is denoted by ``$+1$" in Table~\ref{classificationTable};
and (iii) $\mathcal{H} ( {\bf k} )$ is symmetric under TRS (or PHS) and $T$ (or $C$) squares to $-1$, which is denoted by ``$-1$" in Table~\ref{classificationTable}.
Hence, there are $3 \times 3 = 9$ possibilities for how a Hamiltonian can transform under both TRS and PHS. 
For eight of these cases, the presence or absence of SLS is fully determined by how $\mathcal{H} ({\bf k})$ transforms under TRS and PHS. 
However, when both time-reversal and particle-hole symmetry are absent, there exists the extra possibility that the combined symmetry $S \propto T C$ is still present. 
Hence, altogether there are ten cases, that fully exhaust all the possible transformation
properties of a given Hamiltonian $\mathcal{H} ( {\bf k})$ under global symmetries.
These ten cases,
each of which defines a symmetry class,
are listed in the first column of Table~\ref{classificationTable}.
We note that these ten symmetry classes are closely related 
to the ten classes of random matrices introduced by Altland and Zirnbauer~\cite{Zirnbauer:1996fk,altlandZirnbauerPRB10}.

Now, the ten-fold classification of nodal superconductors (and semimetals) does not only depend on the symmetry properties of $\mathcal{H} ( {\bf k} )$ but also 
on the codimension  
\begin{eqnarray}
p= d_{\mathrm{BZ}} - d_{\mathrm{n}}
\end{eqnarray}
of the nodal structure, i.e., on the difference between the dimension $d_{\mathrm{BZ}}$ of the Brillouin zone and the dimension $d_{\mathrm{n}}$ of the superconducting node.
Furthermore, it depends on how the superconducting nodes transform under the nonspatial symmetries~\cite{matsuura2012,chiuSchnyder2014}. Two different situations have to be distinguished: (i) Each point node or line node is left invariant by  nonspatial symmetries and (ii) different point nodes or line nodes are pairwise related
to each other by global symmetry operations. In case (i) the superconducting nodes are located at high-symmetry points of the Brillouin zone, i.e. at time-reversal invariant momenta,
whereas in case (ii) the point or line nodes  are positioned away from high-symmetry points of the Brillouin zone.

\begin{table}[t!]
\caption{Ten-fold classification of nodal points/lines 
and Fermi surfaces in  superconductors and
semimetals, respectively~\cite{matsuura2012,ZhaoPRL13,ZhaoWangPRB14,chiuSchnyder2014}.
The first and second rows indicate   
the codimension  $p=d_{\mathrm{BZ}} - d_{\mathrm{n}}$ of 
the nodal lines at high-symmetry points and away from high-symmetry points of the Brillouin zone, respectively.
The first column lists the ten symmetry classes (using the ``Cartan nomenclature"~\cite{schnyderAIP}), which are distinguished by the presence or absence of TRS ($T$), PHS ($C$), and chiral symmetry ($S$).
The second and third columns indicate 
 the sign of the squared symmetry operators $T^2$ and $C^2$, respectively.
The absence of symmetries is denoted by ``0". The presence of chiral symmetry $S$ is denoted by ``$1$".
}
\label{classificationTable}
\begin{center}
\begin{threeparttable}
\begin{tabular}{|c|ccc|cccccccc|}
\hline
\mbox{\footnotesize{at high-sym.\ point}} & \multirow{2}{*}{$T$} & \multirow{2}{*}{$C$} & \multirow{2}{*}{$S$} & \footnotesize{$p$=8}  & \footnotesize{$p$=1} & \footnotesize{$p$=2} & \footnotesize{$p$=3} & \footnotesize{$p$=4} & \footnotesize{$p$=5} & \footnotesize{$p$=6} & \footnotesize{$p$=7}   \\
\mbox{\footnotesize{off high-sym.\ point}}    &  &  & & \footnotesize{$p$=2} & \footnotesize{$p$=3} & \footnotesize{$p$=4} & \footnotesize{$p$=5} & \footnotesize{$p$=6} & \footnotesize{$p$=7} & \footnotesize{$p$=8} & \footnotesize{$p$=1}  \\
\hline  \hline
     A    & 0 & 0 & 0  &   0 & $\mathbb{Z}$ & 0 & $\mathbb{Z}$ & 0 & $\mathbb{Z}$ & 0    & $\mathbb{Z}$           \\
  AIII    & 0 & 0 & $1$ &  $\mathbb{Z}$ & 0 & $\mathbb{Z}$ & 0 & $\mathbb{Z}$ & 0 & $\mathbb{Z}$  & 0           \\ 
\hline \hline
  AI  & $+1$ & 0 & 0    & 0 & 0 & 0 & $2\mathbb{Z}$ & 0 & $\mathbb{Z}_2^{\footnotesize{\dag,\S}}$ & $\mathbb{Z}_2^{\footnotesize{\dag,\S}}$   &  $\mathbb{Z}$      \\
  BDI & $+1$ & $+1$ & $1$   & $\mathbb{Z}$ & 0 & 0 & 0 & $2\mathbb{Z}$ & 0 & $\mathbb{Z}_2^{\footnotesize{\dag,\S}}$   & $\mathbb{Z}_2^{\footnotesize{\dag,\S}}$      \\
  D & 0 & $+1$ & 0   & $\mathbb{Z}_2{}^{\footnotesize{\dag,\S}}$ & $\mathbb{Z}$ & 0 & 0 & 0 & $2\mathbb{Z}$ & 0  & $\mathbb{Z}_2^{\footnotesize{\dag,\S}}$          \\
  DIII   & $-1$ & $+1$ & $1$   & $\mathbb{Z}_2^{\footnotesize{\dag,\S}}$ & $\mathbb{Z}_2^{\footnotesize{\dag,\S}}$ & $\mathbb{Z}$ & 0 & 0 & 0 & $2\mathbb{Z}$ & 0       \\
  AII  & $-1$ & 0 & 0    & 0 & $\mathbb{Z}_2^{\footnotesize{\dag,\S}}$ & $\mathbb{Z}_2^{\footnotesize{\dag,\S}}$ & $\mathbb{Z}$ & 0 & 0 & 0   & $2\mathbb{Z}$      \\
  CII  & $-1$ & $-1$ & $1$     & $2\mathbb{Z}$ & 0 & $\mathbb{Z}_2^{\footnotesize{\dag,\S}}$ & $\mathbb{Z}_2^{\footnotesize{\dag,\S}}$ & $\mathbb{Z}$ & 0 & 0   & 0     \\
  C & 0 & $-1$ & 0   &   0 & $2\mathbb{Z}$ & 0 & $\mathbb{Z}_2^{\footnotesize{\dag,\S}}$ & $\mathbb{Z}_2^{\footnotesize{\dag,\S}}$ & $\mathbb{Z}$ & 0  & 0         \\
  CI  & $+1$ & $-1$ & $1$   &  0 & 0 & $2\mathbb{Z}$ & 0 & $\mathbb{Z}_2^{\footnotesize{\dag,\S}}$ & $\mathbb{Z}_2^{\footnotesize{\dag,\S}}$ & $\mathbb{Z}$   & 0 \\
\hline \hline
\end{tabular}
\begin{tablenotes}
\item[${}^{\textrm{$\dag$}}$]  $\mathbb{Z}_2$ numbers only protect nodal lines of dimension zero ($d_{\mathrm{n}}=0$) at high-symmetry points of the Brillouin zone.
\item[${}^{\textrm{$\S$}}$] Nodal lines of any dimension $d_{\mathrm{n}}$ located away from high symmetry points of the Brillouin zone cannot be protected by a $\mathbb{Z}_2$ number. 
However, the $\mathbb{Z}_2$ topological invariant can lead
to the appearance of protected zero-energy surface states at time-reversal invariant momenta of the surface Brillouin zone.
    \end{tablenotes}
    \end{threeparttable}
\end{center}
\end{table}

\subsection{Nodes at high-symmetry points}

Let us start by discussing the ten-fold classification of superconducting nodes located at high symmetry points of the Brillouin zone~\cite{matsuura2012,ZhaoPRL13,ZhaoWangPRB14,shiozakiPRB14,chiuSchnyder2014}.
For this case the classification of point nodes can be derived from the ten-fold classification of fully gapped superconductors trough a dimensional reduction procedure~\cite{matsuura2012}. 
That is, the surface state of a fully gapped $d_{\mathrm{BZ}}$-dimensional topological superconductor realizes a topologically stable point node in $d_{\mathrm{BZ}}-1$ dimensions.
Indeed, the bulk topological invariant of the $d_{\mathrm{BZ}}$-dimensional fully gapped superconductor is directly related  
to the topological charge of the nodal point at the boundary~\cite{shiozakiPRB14,essinGurariePRB11}.
From these observations it follows that the classification of global-symmetry-invariant nodal points with $d_n = 0$ is obtained from  the original ten-fold classification of 
fully gapped superconductors by the dimensional shift $d_{\mathrm{BZ}} \to d_{\mathrm{BZ}} - 1$.
This reasoning also holds for nodal structures with codimension $p < d_{\mathrm{BZ}}$ (i.e., $d_{\mathrm{n}} > 0$), provided that they are protected by 
a $\mathbb{Z}$ invariant or  $2\mathbb{Z}$ invariant. $\mathbb{Z}_2$ numbers, on the other hand, guarantee only the stability of nodes with $d_{\mathrm{n}} = 0$, i.e., point nodes.
These findings are  confirmed by derivations based on K theory~\cite{ZhaoPRL13,ZhaoWangPRB14,shiozakiPRB14} and minimal Dirac Hamiltonians~\cite{chiuSchnyder2014}.
The classification of global-symmetry-invariant nodal structures (and Fermi surfaces) is 
summarized in Table~\ref{classificationTable}, where the first row indicates the codimension $p$ of the superconducting nodes.
For any codimension $p$ there are three symmetry classes for which stable superconducting nodes (or Fermi surfaces) exist that are protected by a
$\mathbb{Z}$ invariant or  $2\mathbb{Z}$ invariant, where the prefix ``$2$" indicates that the topological number only takes on even values.
Furthermore, in each spatial dimension $d_{\mathrm{BZ}}$ there exist two symmetry classes that 
allow for topologically stable point nodes (Fermi points) which are protected by a binary $\mathbb{Z}_2$ number.

\subsection{Nodes off high-symmetry points}
\label{secClassOff}

Second, we review the topological classification of superconductors with nodes that are located away from high-symmetry points of the Brillouin zone~\cite{matsuura2012,chiuSchnyder2014}.
These point or line nodes are pairwise mapped onto each other by the global antiunitary symmetries, which relate ${\bf k}$ to $-{\bf k}$.
An analysis based on the minimal-Dirac-Hamiltonian method~\cite{chiuSchnyder2014} shows that 
only $\mathbb{Z}$-type invariants can guarantee the stability of superconducting nodes off high-symmetry points, whereas
 $\mathbb{Z}_2$ numbers do not lead to stable nodes. However, as illustrated in terms of the example of Sec.~\ref{SecExampNCS}, $\mathbb{Z}_2$ numbers
 may nevertheless give rise to zero-energy surface states at time-reversal-invariant momenta of the surface Brillouin zone. 
The complete classification of superconducting nodes that are located away from high-symmetry points is presented in
Table~\ref{classificationTable}, where the second row gives the codimension $p$ of the superconducting node.
Observe that this classification scheme is related to the ten-fold classification of fully gapped superconductors and insulators by
the dimensional shift $d_{\mathrm{BZ}} \to d_{\mathrm{BZ}} + 1$.

\subsection{Examples}
\label{SecTopoExamp}

\begin{figure}
\begin{center}
\includegraphics[clip,width=1.0\columnwidth]{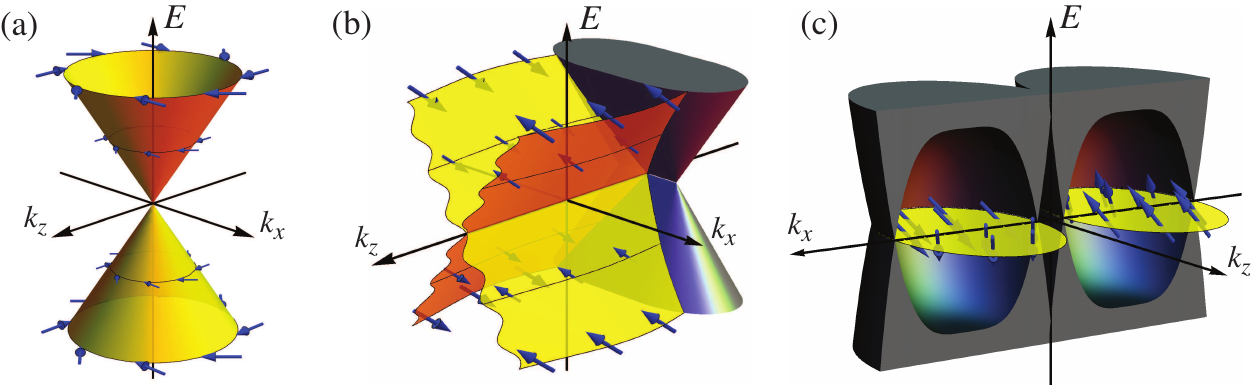}
\end{center}
  \caption{\label{surfStates} 
Energy spectrum of the three different types of topological subgap states that can exist at the surface of nodal noncentrosymmetric superconductors:
(a) helical Majorana cone, (b) arc surface state, and (c)  flat-band surface state.  
Figure adapted from Ref.~\cite{hofmannPRB13}.}   
\end{figure}

For the phenomenological model Hamiltonians given in Sec.~\ref{secModels}, we derive in this subsection explicit expressions for the topological invariants
that protect the superconducting nodes against gap opening.
We also use these  examples to illustrate the bulk-boundary correspondence~\cite{essinGurariePRB11,grafCommMatPhys13}, which links the topological characteristics of the nodal gap structure to the
appearance of zero-energy states at the boundary.  Depending on the case, these zero-energy surface states are either linearly dispersing Majorana cones, Majorana
flat bands, or arc surface states (cf.~Fig.~\ref{surfStates}). We note that in real superconducting materials the gap nodes
are usually positioned away from the high-symmetry points of the Brillouin zone.
Indeed, this is the case for the three examples of Sec.~\ref{secModels}, which are therefore classified
according to Sec.~\ref{secClassOff}.

%%%%%%%%%%%%%%%%%%%%%%%%%%%%
%%%%%%%%%%%%%%%%%%%%%%%%%%%%

\subsubsection{The A phase of ${}^3$He}
\label{SecExampChiralP}

The A phase of ${}^3$He is phenomenologically described by
Hamiltonian~\eqref{hamAphase}, which 
satisfies particle-hole symmetry $C^{-1} \mathcal{H} (-{\bf k} ) C 
=  - \mathcal{H} ( {\bf k} )$ with $C=  \mathcal{K} \tau_x $. Time-reversal symmetry, however, is broken, because the superconducting order parameter of Eq.~\eqref{hamAphase} is complex.
Hence, since $C^2 = + \mathbbm{1}$, Hamiltonian~\eqref{hamAphase} belongs to symmetry class D. 
We infer from  Table~\ref{classificationTable} that the 
Weyl nodes of the Hamiltonian~\eqref{hamAphase}, which have codimension
$p=3$ and occur off high symmetry points, are protected by a
$\mathbbm{Z}$ topological number.
In order to derive a formula for this topological number 
it is convenient to rewrite Hamiltonian~\eqref{hamAphase} as 
\begin{eqnarray}
\mathcal{H} ( {\bf k} )
= 
{\bf N} ( {\bf k} ) \cdot {\bm \tau},
\end{eqnarray}
i.e., a dot product between the pseudospin vector ${\bf N} ({\bf k} )
= \left( h ( {\bf k}), \Delta_0  k_x / k_F  , \Delta_0  k_y / k_F
\right)$  and the vector of Pauli matrices ${\bm \tau} = ( \tau_x,
\tau_y, \tau_z )$. 
The unit vector field ${\bf n}_{\bf k} = {\bf N} ( {\bf k} ) / | {\bf N} ( {\bf k} ) |$
exhibits singular points at the Weyl nodes ${\bf K}_{\pm}$ of $\mathcal{H} ({\bf k})$. These point singularities realize (anti)hedgehog  defects in momentum space and are characterized by
the Chern number~\cite{Volovik:book,VolovikExotic,volovikLectNotes13,Volovik3HeA}
\begin{eqnarray} \label{ChernAphase}
N_{\mathcal{C}} 
=
\frac{1}{4 \pi }
\int_{\mathcal{C} } 
d^2  \widetilde{\bf k}  \; 
{\bf n}_{\bf k}  \cdot \left[ \partial_{k_1} {\bf n}_{\bf k}  \times \partial_{k_2} {\bf n}_{\bf k}   \right] ,
\end{eqnarray}
where $\mathcal{C}$ denotes a two-dimensional surface parametrized by $\widetilde{\bf k} = (k_1, k_2)$, which encloses one of the two Weyl nodes. 
Choosing $\mathcal{C}$ to be a sphere $S^2$
centered at one of the two Weyl points, 
the integral~\eqref{ChernAphase} can be straightforwardly
evaluated. The nodes 
at ${\bf K}_{\pm} = ( 0,  0, \pm k_F)$  have topological charge $N_{S^2} = \pm 1$. 
We note that ${\bf n}_{\bf k}$ restricted to the sphere $S^2$ in momentum space defines a map from ${\bf k} \in S^2$ to the space
${\bf n}_{\bf k} \in S^2$. 
The Chern number $N_{S^2}$, Eq.~\eqref{ChernAphase}, distinguishes between different homotopy classes of these maps~\cite{Nakahara:2003ve}.
Since the homotopy group $\pi_2 ( S^2 )$ equals the group of the integer numbers $\mathbb{Z} $, there 
 is an infinite number of homotopy classes of  maps ${\bf k} \in S^2 \mapsto  {\bf n}_{\bf k}$ 
and $N_{S^2}$ can in general take on any integer value.

\begin{figure}
\begin{center}
\includegraphics[clip,width=0.7\columnwidth]{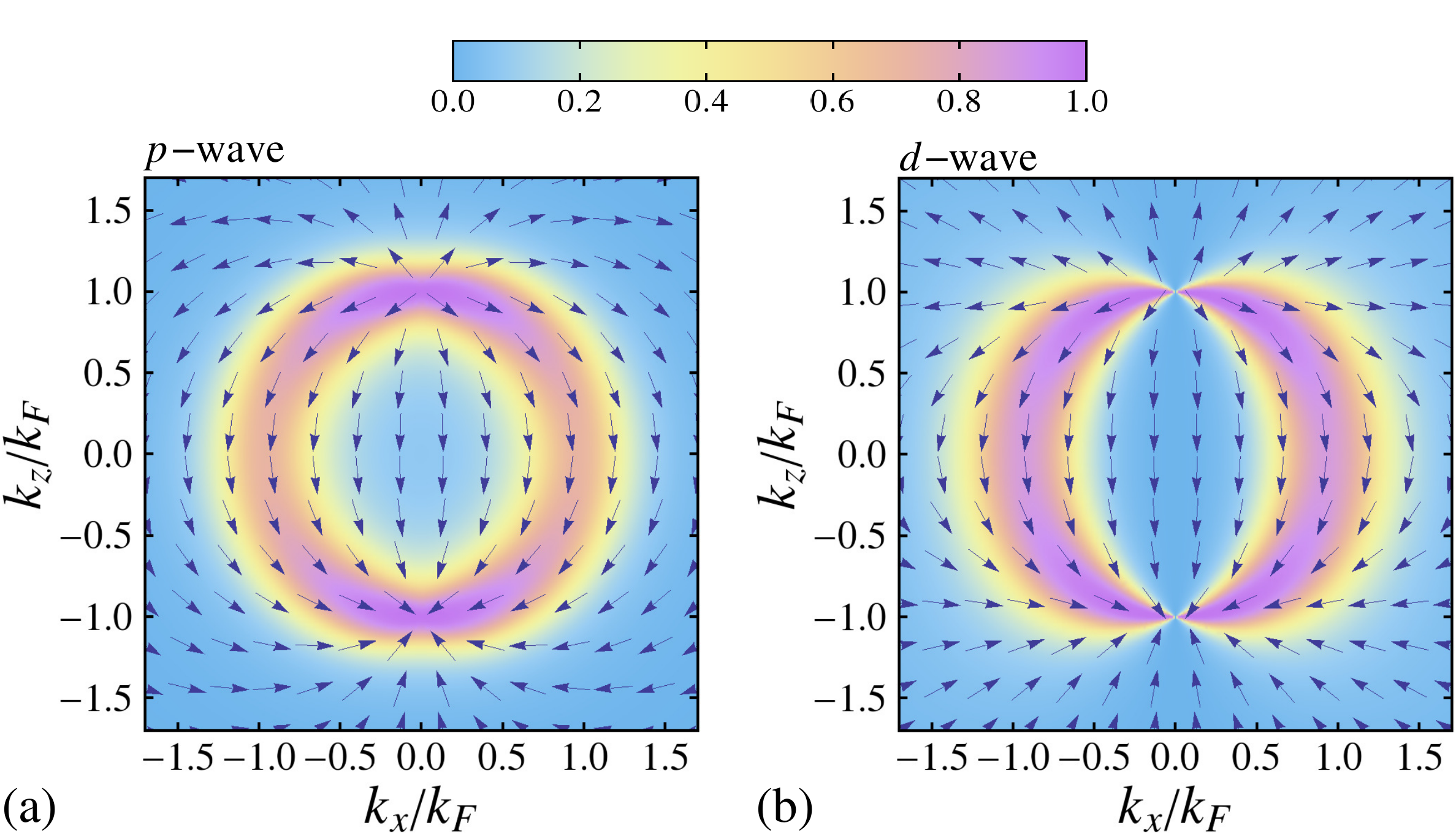}
\end{center}
  \caption{\label{Berry} 
Berry curvature in (a) the ($p_x+ip_y$)-wave and (b) the
($d_{x^2-y^2}+id_{xy}$)-wave Weyl superconductor in the $k_x$-$k_z$
plane. The arrows indicate the direction of ${\bf F}({\bf k})$ while
the colour scale gives $\frac{2}{\pi}\arctan(|{\bf F}({\bf k})|)$. We
choose $\Delta_0 = 0.5\mu$; realistic values of $\Delta_0\ll\mu$ give
a curvature which is much more sharply peaked at the Fermi
surface.}    
\end{figure}

Alternatively, instead of considering a sphere $S^2$ one can take
$\mathcal{C}$ to be a plane perpendicular to the $k_z$ axis in the
Brillouin zone. This yields $N_{k_z} = +1$   for  $ | k_z | <  k_F$
and zero otherwise.  That is, the unit vector ${\bf n}_{\bf k}$ for
fixed $| k_z | < k_F$ possesses a unit skyrmion texture in momentum
space and $N_{k_z}$ measures its skyrmion number. In this picture, we
can give the topological invariant~\eqref{ChernAphase} a more physical
meaning by reexpressing it in terms of the Berry curvature
$ {\bf F} ( {\bf k}  )
=
{\bm \nabla}_{\bf k}  \times {\bf A} ( {\bf k} )$, i.e.,
\begin{eqnarray} \label{eq:Chernkz}
N_{k_z} 
&=&
\frac{1}{2 \pi } \int d k_x d k_y \, F_z ( {\bf k} ).
\end{eqnarray}
 Here, 
${\bf A} ( {\bf k} ) = i \left\langle u_- ( {\bf k}) \right|  {\bm \nabla}_{\bf k}  \left| u_- ({\bf k} ) \right\rangle$
denotes the Berry connection (also known as Berry vector potential), 
which is obtained from the negative-energy wavefunctions $u_- ({\bf k} )$ of the BdG Hamiltonian.
For Hamiltonian \eqref{hamAphase}, the three components 
of the Berry curvature are given by
\begin{eqnarray} \label{berryCurvatureChiralPwave}
F_x ( {\bf k} )
= 
\frac{ \Delta^2_0 k_F^2 k_x k_z }
{  \mu^2 \left[  ( k^2 - k_F^2 )^2 + \frac{\Delta^2_0}{ \mu^2}  k_F^2 ( k_x^2 + k_y^2)  \right]^{\frac{3}{2} }  }  ,
\nonumber\\
F_y ( {\bf k} )
=
\frac{ \Delta^2_0 k_F^2 k_y k_z }
{  \mu^2 \left[  ( k^2 - k_F^2 )^2 + \frac{\Delta^2_0}{ \mu^2}  k_F^2 ( k_x^2 + k_y^2)  \right]^{\frac{3}{2} }  }  ,
\\
F_z ( {\bf k} )
=
\frac{ \Delta_0^2  k_F^2 ( k_z^2 - k_x^2 - k_y^2 - k_F^2 )  }
{2 \mu^2 \left[ (k^2 -k_F^2 )^2 + \frac{\Delta^2_0}{ \mu^2} k_F^2 (k_x^2 + k_y^2)   \right]^{\frac{3}{2} } } .
\nonumber
\end{eqnarray} 
The Berry curvature is plotted in~\fig{Berry}(a).
We observe from Eq.~\eqref{berryCurvatureChiralPwave}  that the Weyl nodes are sources
and drains of Berry flux. That is, they act as unit (anti-)monopoles of the Berry curvature ${\bf F} ( {\bf k} )$
and there is a Berry flux of $2 \pi$ flowing from one Weyl node to the other along the $z$ direction.
Along the $x$ and $y$ directions,
on the other hand, the Berry flux is vanishing since $F_{x (y)} ( {\bf k}) $ are
odd in $k_x$ ($k_y$) and $k_z$. We also note that the Berry
curvature is sharply peaked within a shell of width $~\Delta_0$
about the Fermi surface.

Due to the bulk-boundary correspondence, the band topology of Weyl
superconductors gives rise to arc surface states. Consider for example
Hamiltonian~\eqref{hamAphase}  in a three-dimensional slab geometry
with a (100) surface. Because of translation invariance along the
surface, the three-dimensional (100) slab can be decomposed into a
family of two-dimensional (10) ribbons parametrized by $k_z$. Ribbons
with a $k_z$ in between the pair of Weyl nodes have a non-zero Chern
number, i.e., $N_{k_z} = +1$ for $ | k_z | < k_F$, 
while for $| k_z | > k_F$ the topology is trivial.
Each ribbon with $ | k_z | < k_F$ can be interpreted as a two-dimensional fully gapped $(p_x + i p_y)$-wave superconductor with a chiral Majorana edge mode. Hence, the Majorana surface states of the Weyl superconductor form a one-dimensional open arc
in the surface Brillouin zone, connecting the projected Weyl nodes of the bulk gap, see Fig.~\ref{Weylsurface}(a) and Sec.~\ref{secBSWeylSC}.
These chiral surface states give rise to anomalous spin and thermal Hall
effects which is proportional to the separation of the Weyl nodes in
momentum space, see Sec.~\ref{SecSurfaceCurrents}.

%%%%%%%%%%%%%%%%%%%%%%%%%%%%
%%%%%%%%%%%%%%%%%%%%%%%%%%%%

\subsubsection{Chiral ($d_{x^2-y^2} \pm i d_{xy}$)-wave superconductor}
\label{SecExampChiralD}

The spin-singlet chiral $(d_{x^2-y^2} \pm i d_{xy}$)-wave superconductor~\eqref{hamChiralD}
satisfies both particle-hole symmetry and $SU(2)$ spin-rotation symmetry  
and therefore belongs to symmetry  class C in Table~\ref{classificationTable}
(or alternatively to class A if one considers only a $U(1)$ part of
the spin-rotation symmetry). Hence, as indicated by the classification table,
the Weyl nodes of Hamiltonian~\eqref{hamChiralD} (which occur off high
symmetry points) are protected by 
a $2 \mathbb{Z}$ invariant, which takes on only even values.
This topological number is given by the same expression as Eq.~\eqref{ChernAphase}  but
with ${\bf N} ( {\bf k} ) =  \left( h ( {\bf k}), \Delta_0  ( k^2_x - k^2_y ) / k^2_F  ,  2 \Delta_0  k_x k_y / k^2_F  \right)$.
As opposed to the previous example, the unit vector ${\bf n}_{\bf k} = {\bf N} ( {\bf k} ) / | {\bf N} ( {\bf k} ) |$
for fixed  $|k_z| < k_F$ now describes a double Skyrmion texture with $N_{k_z} = \pm 2$.
That is, the double Weyl points at ${\bf K}_{\pm} = (0,0, \pm k_F)$
correspond  to double (anti-)monopoles of the Berry curvature ${\bf F}
( {\bf k} )$, which is given by~\cite{goswamiArXiv13} 
\begin{eqnarray}
F_x ( {\bf k} )
= 
\frac{2 \Delta_0^2  k_x k_z ( k_x^2 + k_y^2) }
{\mu^2  \left[  ( k^2 - k_F^2 )^2 + \frac{ \Delta_0^2}{\mu^2} (k_x^2 + k_y^2)^2   \right]^{\frac{3}{2}}} ,
\nonumber\\
F_y ( {\bf k} )
=
\frac{2 \Delta_0^2  k_y k_z ( k_x^2 + k_y^2) }
{\mu^2  \left[  ( k^2 - k_F^2 )^2 + \frac{ \Delta_0^2}{\mu^2} (k_x^2 + k_y^2)^2   \right]^{\frac{3}{2}}} ,
\\
F_z ( {\bf k} )
=
\frac{ 2 \Delta_0^2 ( k_z^2 - k_F^2 ) (k_x^2 + k_y^2 ) }
{\mu^2 \left[ (k^2 -k_F^2 )^2 + \frac{\Delta_0^2}{ \mu^2}Ê( k_x^2 + k_y^2)^2  \right]^{\frac{3}{2}}   } .
\nonumber
\end{eqnarray}
The Berry curvature is plotted in~\fig{Berry}(b). 
As a result of the topological non-triviality, the $d$-wave Weyl
superconductor~\eqref{hamChiralD} supports two spin-degenerate,
chirally dispersing arc surface states, see
Fig.~\ref{Weylsurface}(b). These surface states carry anomalous spin
and thermal Hall currents.

\subsubsection{Nodal noncentrosymmetric superconductor}
\label{SecExampNCS}

Nodal NCSs defined by the BdG
Hamiltonian $\mathcal{H} ( {\bf k} )$ in Eq.~\eqref{NCSham}
satisfy time-reversal symmetry $T^{-1} \mathcal{H} ( - {\bf k} ) T = + \mathcal{H} ( {\bf k} )$, with $T = \mathcal{K}  \tau_0 \otimes i \sigma_y $, and particle-hole 
symmetry $C^{-1} \mathcal{H} (- {\bf k} ) C = - \mathcal{H} ( {\bf k} )$, with $C = \mathcal{K}  \tau_x \otimes \sigma_0$. 
Here, $\tau_i$ and $\sigma_i$ are the Pauli matrices operating in particle-hole and spin space, respectively.
Since $T^{2} = -  \mathbbm{1}$ and $C^2 = +  \mathbbm{1}$, Hamiltonian~\eqref{NCSham} belongs to symmetry class DIII.
According to Table~\ref{classificationTable}, the topological properties  of class DIII Hamiltonians
are characterized by $\mathbb{Z}_2$ topological numbers.
While these binary invariants can give
rise to protected Majorana cone surface states or arc surface states,
they do not guarantee the stability of the superconducting nodes~\cite{chiuSchnyder2014}.
Hence, one might conclude that the superconducting nodes of the NCS~\eqref{NCSham} are unstable against gap opening.
In addition to TRS and PHS, however, $\mathcal{H} ( {\bf k} )$ in
Eq.~\eqref{NCSham} also possesses the chiral symmetry
$S^{-1} \mathcal{H} ( {\bf k} ) S = - \mathcal{H} ({\bf k} )$, with $S= - \tau_x \otimes \sigma_y$.
Therefore, $\mathcal{H} ( {\bf k} )$ can be viewed as a member of symmetry class AIII and, as indicated by Table~\ref{classificationTable}, its line nodes are protected by a $\mathbb{Z}$ invariant, i.e., a winding number. 
 
The $\mathbb{Z}_2$ and $\mathbb{Z}$ invariants of symmetry class DIII and AIII, respectively, can be conveniently expressed in terms
of the off-diagonal block  $q ( {\bf k} ) $ of the spectral projector.
For Hamiltonian~\eqref{NCSham}, the matrix $q ( {\bf k} ) $ takes the following form~\cite{Schnyder10,Schnyder12}
\begin{eqnarray} \label{Qmatrix}
\fl \quad
q ( {\bf k} ) =
\frac{1}{2 \Lambda_{1, {\bf k}} \Lambda_{2, {\bf k}}}
\left[
\left(
\Lambda^+_{\bf k} B^{\phantom{+}}_{\bf k} - \Lambda^-_{\bf k}  A^{\phantom{+}}_{\bf k} \left| {\bf l}_{\bf k} \right|Ê
\right) \sigma_0
+
\left(
\Lambda^+_{\bf k} A^{\phantom{+}}_{\bf k}Ê\left| {\bf l}_{\bf k} \right|  - \Lambda^-_{\bf k} B^{\phantom{+}}_{\bf k}
\right) \frac{ {\bf l}_{\bf k} }{ \left| {\bf l}_{\bf k} \right| } \cdot {\bm \sigma}
\right] .
\end{eqnarray}
In Eq.~\eqref{Qmatrix} we have introduced the short-hand notation
$A_{\bf k} = \lambda + i \Delta_t $, $B_{\bf k} = \varepsilon_{\bf k} + i \Delta_s $,
and $\Lambda^{\pm}_{\bf k}  = \Lambda_{1, {\bf k} } \pm \Lambda_{2, {\bf k} }$, where
\begin{eqnarray}
\Lambda_{1, {\bf k} } = \sqrt{ \left( \xi^+_{\bf k} \right)^2 + \left( \Delta^+_{\bf k} \right)^2 }
\quad \textrm{and} \quad
\Lambda_{2, {\bf k} } = \sqrt{ \left( \xi^-_{\bf k} \right)^2 + \left( \Delta^-_{\bf k} \right)^2 }
\end{eqnarray}
are the positive eigenvalues of $\mathcal{H} ( {\bf k} )$.
The unitary matrix $q ( {\bf k} ) \in U(2) $  is constrained by time-reversal symmetry as $ i \sigma_y q^{\mathrm{T}} ( - {\bf k}Ê) = q ( {\bf k}Ê) i \sigma_y$.

\paragraph{Winding number}

The stability of the line nodes of NCS~\eqref{NCSham} is ensured by the conservation of the winding number~\cite{beriPRB10,Schnyder10,matsuura2012,Sato11,tanakaJPSJ2012}
\begin{equation} \label{1DwindingNo}
W_{\mathcal{L}}
=\frac{1}{2\pi i} \oint_\mathcal{L} dk_l \, \mathrm{Tr} [q^{\dag} ( {\bf k} ) \partial_{k_l } q ( {\bf k} ) ],
\end{equation}
where $\mathcal{L}$ is a closed one-dimensional contour parametrized by $k_l$, which interlinks with a line node.
Mathematically speaking, Eq.~\eqref{1DwindingNo} labels the homotopy classes of mappings from $\mathcal{L} \simeq S^1 \to 
q ( {\bf k} ) \in U(2)$. Since the first homotopy group of $U(2)$ equals the group
 of the integer numbers $\mathbb{Z}$ \cite{Nakahara:2003ve}, i.e., $\pi_1 [ U (2) ] = \mathbb{Z}$, the winding number $W_{\mathcal{L}}$ can in principle
 take on any integer value.
Using Eq.~\eqref{Qmatrix}, we find that $W_{\mathcal{L}}$ for
the  BdG Hamiltonian~\eqref{NCSham} 
can be rewritten in the simpler form 
\begin{eqnarray} \label{windingNo2}
W_{\mathcal{L}} =
\frac{1}{2 \pi }
\oint_{\mathcal{L}}
d k_l  \, \partial_{k_l}
\left[
\mathrm{arg} \left( \xi^+_{\bf k} + i \Delta^+_{\bf k} \right) + \mathrm{arg} \left( \xi^-_{\bf k} + i \Delta^-_{\bf k} \right)
\right].
\end{eqnarray} 
Assuming that the superconducting gaps  $\Delta^{\pm}_{\bf k}$ of Hamiltonian~\eqref{NCSham}
are nonzero only in the vicinity of the Fermi surfaces (which is the case for weak-pairing superconductors),
Eq.~\eqref{windingNo2} can be further simplified to~\cite{Schnyder12}  
\begin{eqnarray} \label{windingNo3}
W_{\mathcal{L}} =
- \frac{1}{2} \sum_{ \nu = \pm } \, \sum_{ {\bf k}_0  \in S^\nu_{\mathcal{L}} }
\mathrm{sgn} \left( \left. \partial_{ k_l} \xi^{\nu }_{\bf k}  \right|_{ {\bf k} = {\bf k}_0 } \right) \mathrm{sgn} \left( \Delta^{\nu}_{ {\bf k}_0}  \right)  .
\end{eqnarray}
The second sum in Eq.~\eqref{windingNo3} is over the set of points $S^\nu_{\mathcal{L}}$
 given by the intersection of the helicity-$\nu$ Fermi surface with the one-dimensional contour $\mathcal{L}$. 
Eq.~\eqref{windingNo3} demonstrates that for a weak-pairing nodal superconductor of the from~\eqref{NCSham},
the topological characteristics of the nodal structure is entirely determined by the phases of the gaps in the neighborhood
of the positive- and negative-helicity Fermi surfaces. This is in complete agreement with the discussion of Sec.~\ref{secBoundStatesNCS}.

\begin{figure}
\begin{center}
\includegraphics[clip,width=0.67\columnwidth]{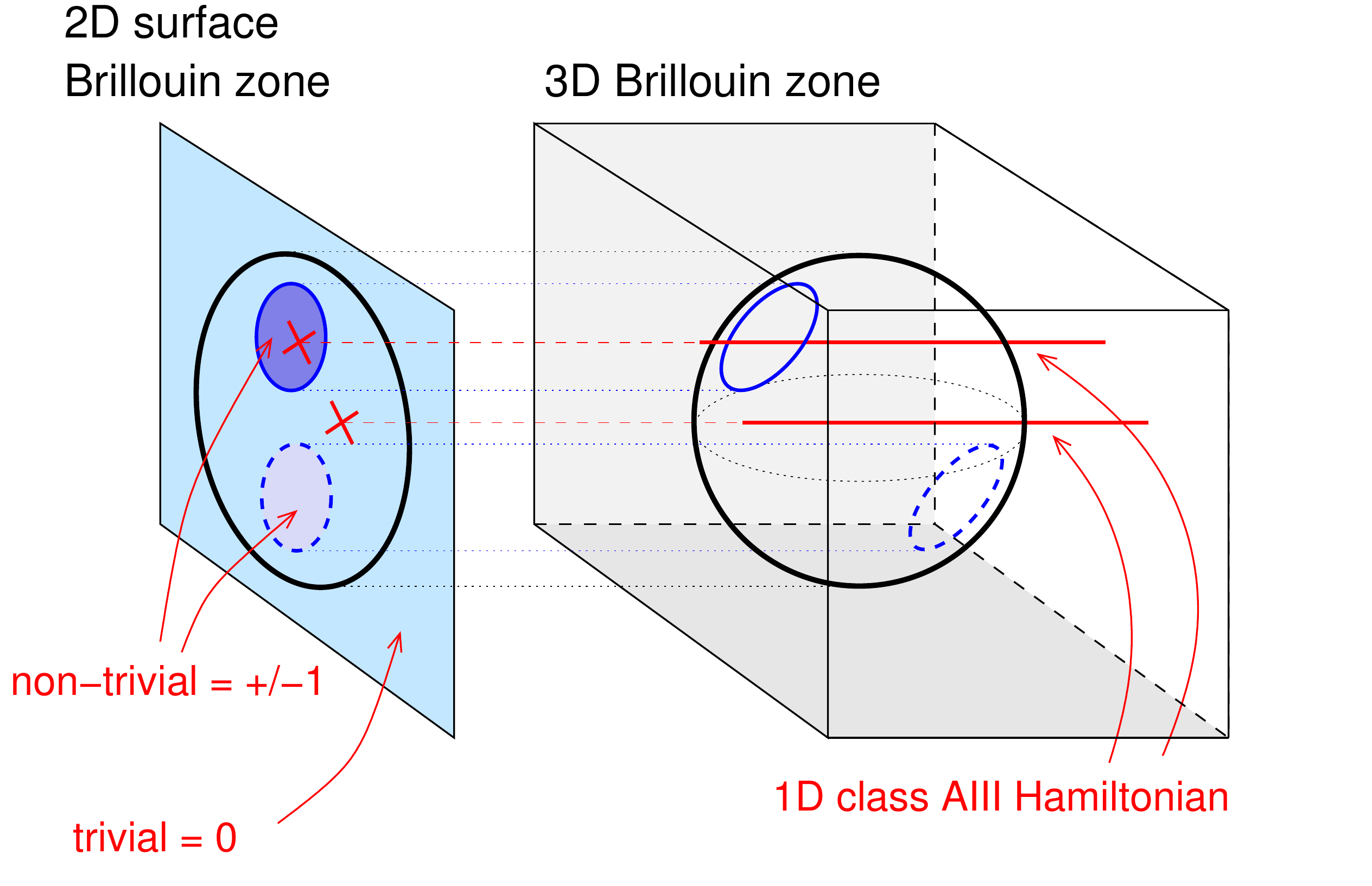}
\end{center}  
  \caption{\label{Figprojection} The relationship of the bulk gap structure to 
the surface states of a nodal topological superconductor. The left
part of the figure shows the surface Brillouin zone with the projected
Fermi surface indicated in black. Flat-band surface states occur
within the two regions bounded by the projected nodal lines (dark blue
and light gray areas). Within these two regions the winding
number~\eqref{windingNoSurf} takes on the values $W = \pm 1$, while
outside these regions it is zero; the bulk Hamiltonians restricted to
a surface momentum in these regions (red lines in 3D Brillouin zone)
belong to symmetry class AIII and are topologically trivial
and nontrivial, respectively. The right part of the figure shows
the three-dimensional bulk Brillouin zone with a spherical Fermi
surface (black ellipse) and two nodal rings (solid and dashed blue ellipses).
}   
\end{figure}

By the bulk-boundary correspondence~\cite{essinGurariePRB11}, a non-zero value of $W_{\mathcal{L}}$ signals the appearance of 
zero-energy states at the surface of the NCS.
To make this more explicit, let us consider a contour $\mathcal{L}$
along a line parallel to the $(lmn)$ direction 
in the bulk Brillouin zone, see Fig.~\ref{Figprojection}. 
With this choice we obtain for Eq.~\eqref{windingNo3}
\begin{eqnarray} \label{windingNoSurf}
W_{(lmn)} ( {\bf k}_{\parallel} )
=
- \frac{1}{2} \sum_{\nu = \pm} \left[ \mathrm{sgn}
\left(  \Delta^{\nu}_{ {\bf k}_{\nu } }  \right)
- \mathrm{sgn} \left(
\Delta^{\nu}_{ {\bf k}^{\prime}_{\nu } }
\right)
 \right] , 
\end{eqnarray} 
where ${\bf k}_{\parallel} $ denotes the momentum perpendicular to the $(lmn)$ direction, and
${\bf k}_{\pm} = ( {\bf k}_{\parallel}, k_{\perp, \pm} )  $  and ${\bf k}^{\prime}_{\pm} =  ( {\bf k}_{\parallel}, k^{\prime}_{\perp, \pm} ) $
are momenta on the positive- and negative-helicity Fermi surfaces which
project onto ${\bf k}_{\parallel}$, where $k_{\perp , \pm}$ and
$k^{\prime}_{\perp, \pm}$ have opposite signs. 
With this, it follows from the bulk-boundary correspondence that zero-energy states
appear at the $(lmn)$ face of the NCS whenever $W_{(lmn)}
({\bf k}_{\parallel})\ne 0$.
This  corresponds to regions of the surface Brillouin zone that are bounded by the projected nodal rings of the bulk gap, see Fig.~\ref{Figprojection}. In other words, 
the non-trivial topology of the nodal rings in the bulk leads to protected flat-band surface states (cf.~Fig.~\ref{surfStates}(c) and Sec.~\ref{secFlatBandQuasiClassics}).

%%%%%%%%%%%%%%%%%%%%%%%%%%%%
%%%%%%%%%%%%%%%%%%%%%%%%%%%%

 \paragraph{$\mathbb{Z}_2$ topological invariant}

The one-dimensional (two-dimensional) $\mathbb{Z}_2$ invariant of
symmetry class DIII is given
by~\cite{ryuNJP2010,Schnyder10,satoPRB06,satoPRB09,qiZhang2010} 
\begin{eqnarray} \label{Z2invariant}
N_{\mathcal{L}}
&=&
\prod_{{\bf K} \in \mathcal{L} }
\frac{\mathrm{Pf}\, \left[ i \sigma_y q^{\mathrm{T}}({\bf K})\right] }
{\sqrt{ \det \left[ i \sigma_y q^{\mathrm{T}} ( {\bf K} ) \right] }},
\end{eqnarray}
where $\mathrm{Pf}$ is the Pfaffian  and the product is over the two (four) time-reversal invariant momenta ${\bf K}$ of
the one-dimensional (two-dimensional) centrosymmetric contour $\mathcal{L}$.
In order for $N_{\mathcal{L}}$ to be well defined, the  contour $\mathcal{L}$ needs to be 
centrosymmetric
(i.e., mapped onto 
itself under ${\bf k} \to - {\bf k}$)
and must not intersect with the nodes of $\mathcal{H} ({\bf k} )$.
Inspection of  Eq.~\eqref{Z2invariant} reveals that $N_{\mathcal{L}}$
can only take on the values $\pm 1$, where $-1$ ($+1$)
indicates a topologically nontrivial (trivial) character.
For weak-pairing superconductors the gaps
 $\Delta^{\pm}_{\bf k}$ are non-zero only
 in a small neighborhood of the Fermi surface and hence Eq.~\eqref{Z2invariant} can be simplified to~\cite{Schnyder12} 
\begin{eqnarray} \label{Z2number}
N_{\mathcal{L}}
=
\mathrm{sgn} \left( \Delta^+_{\mathcal{L}} \right) \mathrm{sgn} \left( \Delta^-_{\mathcal{L}} \right)  ,
\end{eqnarray} 
where $\mathrm{sgn} \left( \Delta^{\pm}_{\mathcal{L}} \right)$ represents the sign of the gap 
at the intersection of $\mathcal{L}$ with the positive/negative helicity Fermi surface.
By the bulk-boundary correspondence \cite{grafCommMatPhys13},
$N_{\mathcal{L}} = -1$ for a one-dimensional  contour $\mathcal{L}$ indicates the appearance of a helical  Majorana cone 
 at a surface of the NCS, see Fig.~\ref{surfStates}(a).
Similarly, for a two-dimensional contour $\mathcal{L}$, $N_{\mathcal{L}} = -1$ gives rise to an arc surface state,
see Fig.~\ref{surfStates}(b).

\subsection{Crystalline topological nodal superconductors}

In the previous subsections we have discussed the topological classification of nodal superconductors in
terms of internal symmetries, i.e., global symmetries that act locally in position space (see Table~\ref{classificationTable}).
Recently it became apparent that besides global symmetries, also crystal symmetries which operate non-locally in 
position space can lead to the protection of superconducting nodes \cite{shiozakiPRB14,chiuSchnyder2014,chiuSchnyderArXiv15,morimotoPRB14,ChangPRB14,YangNagaosaNatComm14}. 
Indeed, a complete topological classification of  reflection-symmetry-protected nodal superconductors
was  recently obtained in Ref.~\cite{chiuSchnyder2014}. 
Moreover, it has been argued that the heavy fermion superconductor UPt$_3$ can be viewed as a crystalline topological nodal superconductor,
since its topological surface states are protected by mirror symmetry~\cite{tsutsumiJPSJ13}.

\section{Bound state wavefunctions}\label{sec:quasiclassical}

In this section we construct the wavefunctions of the
surface states of a topological superconductor within an approximate
low-energy effective theory. Such so-called ``quasiclassical''
methods are a standard technique in the study of inhomogeneous
superconductivity~\cite{SereneRainer1983,Kashiwaya2000,Eschrig2000,EschrigIniotakisTanakabook}. The
focus here is to demonstrate  
how this approach can be used to explore the surface physics of
topological superconductors.

\subsection{Deriving the low-energy theory}

We consider a surface of a topological superconductor,
with normal defining the axis $r_\perp$. The superconductor
is assumed to occupy the half-space $r_\perp>0$, and translational
invariance parallel to the surface ensures that the transverse
momentum ${\bf k}_\parallel$ is a good quantum number.
Our aim is to solve the BdG equation
\beq
\tilde{H}\Psi({\bf k}_\parallel; r_\perp) = E\Psi({\bf
  k}_\parallel;r_\perp) \label{eq:BdGeqn}
\eeq
for states with energy comparable to the superconducting gap. The
Hamiltonian $\tilde{H}({\bf k}_{\parallel})$ appearing here is related
to the bulk Hamiltonian $ H({\bf k})$ by
\beq
\tilde{H}({\bf k}_\parallel) = \int \frac{dk_\perp}{2\pi} H({\bf k}) e^{ik_\perp r_\perp} .
\eeq
The BdG equations can be solved numerically, e.g. by exact
diagonalization of a lattice regularization of the
Hamiltonian. Superconductivity is a low-energy phenomenon, however,
where the characteristic energy scale of the gap $\Delta$ is typically
much
smaller than the Fermi energy $E_F$, and it is at such low energy
scales that the surface states exist. Not only is it computationally 
advantageous to restrict our solution of~\eq{eq:BdGeqn} to the low-energy sector, but it also
permits at least semi-analytic solutions which provide
a more transparent understanding of the physics. This is the
motivation for the quasiclassical theory of superconductivity. A full
development of this theory, in particular the construction of the
quasiclassical Green's function and the self-consistent
determination of the pairing potential, is rather involved and beyond
the scope of this review. We therefore sketch the theory 
below, and refer interested readers to the
extensive literature on the subject, e.g.~\cite{SereneRainer1983,Kashiwaya2000,Eschrig2000,EschrigIniotakisTanakabook}.

The first step in constructing the effective theory is
to bring the 
noninteracting Hamiltonian $h({\bf k})$ into a diagonal form. 
The low-energy
quasiparticle excitations in the superconductor occur
within a narrow momentum shell about the normal-state
Fermi surface of width $\sim \Delta/\hbar|{\bf v}_{F}| \ll |{\bf k}_F|$
where ${\bf v}_F$ is the Fermi velocity. The Fermi velocity is 
regarded as constant across this momentum shell, which allows us to
linearize the low-energy Hamiltonian $[h({\bf
      k})]_{\alpha,\alpha} \approx \hbar{\bf v}_{F,\alpha}\cdot({\bf k} -
  {\bf k}_{F,\alpha})$
where ${\bf k}_{F,\alpha}$ lies on the Fermi surface of the $\alpha$
band. We now examine the pairing matrix 
$\hat{\Delta}({\bf k})$. The superconducting gap is taken to be
constant along the direction of the Fermi velocity at each
point on the Fermi surface, i.e. $\hat{\Delta}({\bf k}) \approx
\hat{\Delta}({\bf k}_{F,\alpha})$. The solution of
the low-energy effective theory is therefore only sensitive to the
value of the superconducting gap in the immediate vicinity of the Fermi
surface. Pairing between states on nondegenerate Fermi surfaces is not
considered. 

An appropriate ansatz for an eigenstate of the low-energy
Hamiltonian is
\beq
\Psi({\bf k}_\parallel;{\bf r}) = \sum_{j}a_j \left(\begin{array}{c}
{\bf u}_j \\
{\bf v}_j 
\end{array}
\right)e^{i({\bf k}_{\parallel}\cdot{\bf r}_\parallel + [k_{j} + \delta k_{j}] r_\perp)},  \label{eq:ansatz}
\eeq
where the sum $j$ is over all scattering channels, $a_j$ are
$\mathbb{C}$-number coefficients,
the vector ${\bf k}_j = ({\bf k}_\parallel, k_{j})$ lies
on the Fermi surface, $|\delta k_j|\ll|k_{j}|$ is a small energy-dependent
deviation from the Fermi wavevector ${\bf k}_j$, and the multi-component BCS 
coherence factors ${\bf u}_j$ and ${\bf v}_j$ are functions only of
${\bf k}_j$ and the energy. The coefficients $a_j$ in~\eq{eq:ansatz} are
determined by imposing boundary conditions on the wavefunction at the
surface, which can in principle be derived from the underlying
microscopic theory. This presents the greatest challenge faced by the
low-energy effective theory, as there is no general
prescription of how this should be done, especially for multiband
systems or those with strong spin-orbital mixing. 

In order to avoid this complication, in the following we
restrict our attention to a commonly-used microscopic model of the
normal state where boundary conditions can be explicitly and easily
derived~\cite{schnyderPRB08,Sau_semiconductor_heterostructures,Roman_SC_semi,Schnyder12,Brydon11,dahlhaus2012,huPRL94,tanakaPRL2010,takeiPRB13,Kashiwaya2000,EschrigIniotakisTanakabook,brydonNJP2013,SchnyderPRB10,brydonNJP2015,tanakaNagaosaPRB09,Fujimoto2009,AsanoNCSJJ2011}. Specifically,
we adopt the noninteracting Hamiltonian 
\beq
h({\bf k}) = \left(\frac{\hbar^2}{2m}|{\bf k}|^2 -
\mu\right)\hat{\sigma}_{0} + \lambda {\bf l}_{\bf
  k}\cdot\hat{\pmb{\sigma}} + V_{\mathrm{work}}\Theta(r_\perp)\label{eq:continuumH}
\eeq
This describes a free electron gas, including antisymmetric spin-orbit
coupling of strength $\lambda$ so as to model both centrosymmetric
($\lambda=0$) and noncentrosymmetric ($\lambda\neq0$) systems. In the
latter case, the 
spin-orbit vector ${\bf l}_{\bf k}$ has the following continuum-limit 
representations for the different point groups 
\beqarray
\fl {\bf l}_{\bf k} = \left\{\begin{array}{cc}
( a_1 k_x + a_2 k_y){\bf x}
+ ( a_3 k_x + a_4 k_y){\bf y}
+ a_5 k_z{\bf z}, & C_2 \\
k_y {\bf x} - k_x{\bf y}, & C_{4v} \\
k_x(1 + g_2[k_y^2+k_z^2]){\bf x} + k_y(1 + g_2[k_x^2+k_z^2]){\bf y} +
k_z(1 + g_2[k_x^2+k_y^2]){\bf z}, & {\cal O} \\
k_x(k_y^2-k_z^2){\bf x} + k_y(k_z^2-k_x^2){\bf y} + k_z(k_x^2-k_y^2){\bf z}, & T_{d}
\end{array}\right. 
\eeqarray
The coefficients $a_{j=1,\ldots,5}$ and $g_2$ appearing in the
expressions for 
the $C_2$ and ${\cal O}$ point groups are arbitrary. Our Hamiltonian also
includes a confining potential $V_{\mathrm{work}}$ for the 
surface which we identify with the work function. For our low-energy
theory it is reasonable to take the limit
$V_{\mathrm{work}}\rightarrow\infty$, and so we implicitly
include this potential as the boundary condition that the
wavefunctions vanish at the surface.

\subsection{Bound states of Weyl superconductors}
\label{secBSWeylSC}

\begin{figure}
\begin{center}
\includegraphics[clip,width=0.6\columnwidth]{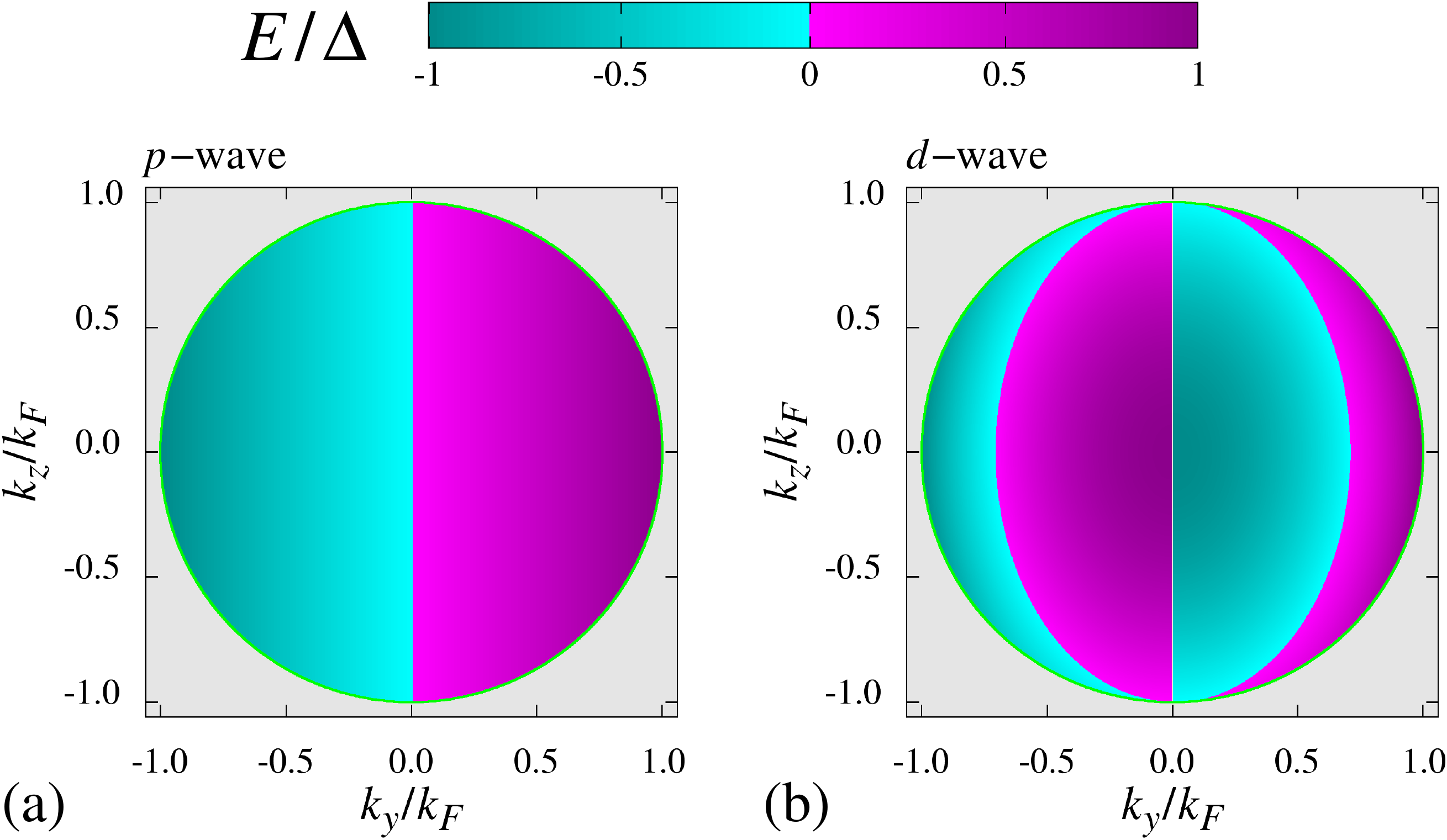}
\end{center}
  \caption{\label{Weylsurface} 
Topological surface states in Weyl superconductors. The
surface bound-state spectra is plotted as a function of the (100) surface 
momentum $(k_y,k_z)$ for (a) the ($p_x+ip_y$)-wave and (b) the
($d_{x^2-y^2}+id_{xy}$)-wave Weyl superconductor. The energy of the
bound states is indicated by the color 
scale, where the jump from cyan to magenta occurs at zero energy;
light grey indicates the absence of 
a surface state. The light green line shows the projection of the
Fermi surface.}    
\end{figure}

As an illustration of the quasiclassical method, we consider the
subgap bound states appearing at the $(100)$ surface of the Weyl
superconductor models introduced in~\Sec{sec:topSCwpointnodes}. Take a
point ${\bf 
  k}_{\parallel} = (k_y,k_z)$ in the  two-dimensional surface
Brillouin zone that lies within the  
projection of the Fermi surface, i.e. $|{\bf k}_\parallel|<k_F$. This
surface wavevector corresponds to two points on the Fermi surface ${\bf
  p}^{}=(k^{}_{x},{\bf k}_\parallel)$ and ${\bf p}^\prime=(-{k}^{}_{x},{\bf
  k}_\parallel)$, where $k^{}_{x} = \sqrt{k_F^2 -
  |{\bf k}_\parallel|^2}$. The bound state wavefunction is then a
superposition of evanescent states at these momenta
\beq
\fl \Psi({\bf k}_\parallel;{\bf r})
 = \left\{a_{{\bf p}} \left(\begin{array}{c}
1\\
\gamma_{{\bf p}}
\end{array}\right)e^{(ik_x - \kappa_{{\bf p}})x} + a_{{\bf p}'} \left(\begin{array}{c}
1\\
\gamma_{{\bf p}'}
\end{array}\right)e^{-(ik_x + \kappa_{{\bf p}'})x}\right\}e^{i{\bf
       k}_\parallel\cdot{\bf r}} \label{eq:Psicase1}
\eeq
where $a_{\bf k}$ are $\mathbb{C}$-number coefficients, and the coherence
factors $\gamma_{\bf k}$ and inverse decay length $\kappa_{{\bf k}}$
are defined  
\beqarray
\gamma_{{\bf k}} &=& \frac{1}{\Delta_{\bf k}}\left[E -
  i \sgn(k_x)\sqrt{|\Delta_{\bf k}|^2 -
    E^2}\right]\,, \\
\kappa_{{\bf k}} &= &\frac{m\sqrt{|\Delta_{{\bf k}}|^2 - E^2}}{\hbar^2|k_{x}|}\,.
\eeqarray
The ansatz~\eq{eq:Psicase1} satisfies the criteria that the
wavefunction vanishes in the bulk. The requirement that the wavefunction
also vanishes at the surface leads to the condition on the coherence factors
\beq
\gamma_{{\bf p}} - \gamma_{{\bf p}'} = 0\,.
\eeq
This can be analytically solved for the surface bound state energies of the
$p$-wave and $d$-wave Weyl superconductor:
\beq
\fl E({\bf k}_\parallel) = 
\left\{\begin{array}{cc}
\Delta_0\left(1 - \tilde{k}_z^2\right)^{1/2}\tilde{k}_y & p_x+ip_y \\
\Delta_0\left(1 - \tilde{k}_z^2\right)\left(1-
  |\tilde{\bf k}_\parallel|^2 - \tilde{k}_y^2\right)\mbox{sgn}(\tilde{k}_y) & d_{x^2-y^2}+id_{xy}
\end{array}\right.
\eeq
where $\tilde{k}_\mu = k_\mu/k_F$.
The bound state spectrum is plotted in~\fig{Weylsurface}. In both
cases, we observe zero-energy arc surface states running between the
nodes at the north and south poles of the Fermi surface. Whereas there
is only a single such arc state in the $p$-wave Weyl superconductor,
there are two arc states in the $d$-wave Weyl superconductor,
consistent with the Chern numbers~\eq{eq:Chernkz} of $N_{k_z} =
1$ and $N_{k_z}=2$ for $|k_z|<k_F$, respectively.

%%%%%%%%%%%%%%%%%%%%%%%%%%%%
%%%%%%%%%%%%%%%%%%%%%%%%%%%%

\subsection{Bound states of NCS}
\label{secBoundStatesNCS}

We now turn our attention to the surface bound states of a NCS. We
restrict ourselves to the limit $E_F \gg \lambda \max\{|{\bf  
  l}_{\bf k}|\} \gg
|\Delta| $, so that we can ignore the spin-splitting of the Fermi
surfaces in the low-energy theory, but the triplet ${\bf d}$-vector
is still constrained to be parallel to the spin-orbit
coupling vector, i.e. ${\bf d}_{\bf k} = \Delta_t {\bf l}_{\bf k}$,
and hence the physical description 
outlined in~\Sec{sec:topSCwlinenodes} is still applicable. This is a common
approximation~\cite{Schnyder12,Brydon11,EschrigIniotakisTanakabook,AsanoNCSJJ2011,luYip10,vorontsovPRL08,iniotakisPRB07,Rahnavard2014,Annunziata2012}
as accounting for the spin-orbit splitting of the Fermi surfaces
complicates the analysis but generally only quantitatively modifies
the surface state spectra. 

%%%%%%%%%%%%%%%%%%%%%%%%%%%%
\begin{figure}
\begin{center}
\includegraphics[clip,width=\columnwidth]{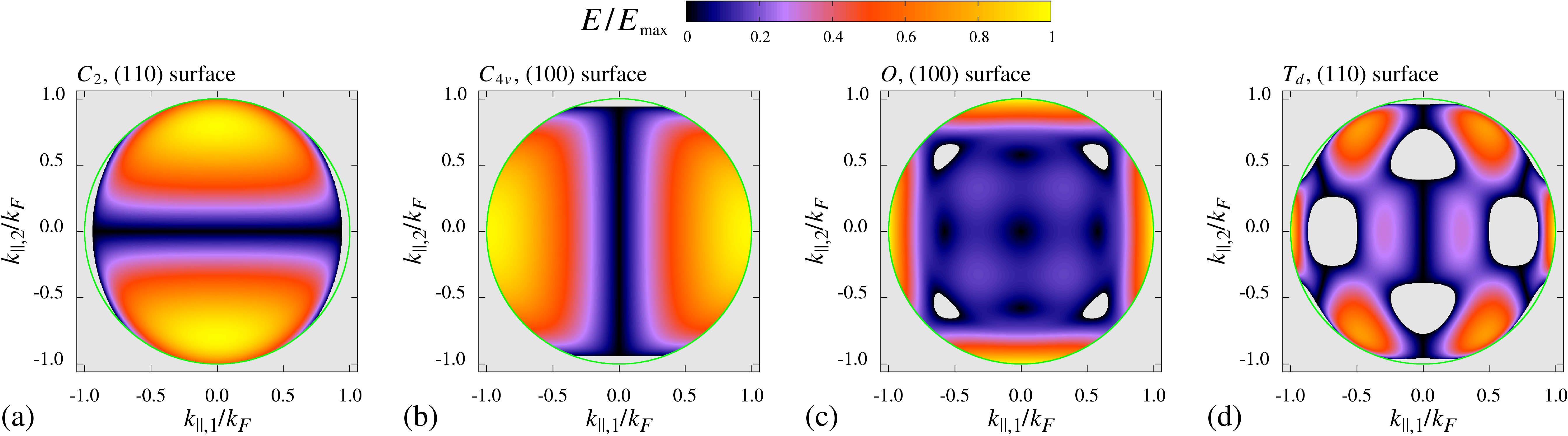}
\end{center}
  \caption{\label{dispersing} 
Helical Majorana and arc states in NCS. The surface
bound-state spectra is plotted as a function of surface 
momentum ${\bf k}_\parallel$ for: (a) (110) face of $C_2$ NCS with $r=0.4$
and $a_{i=1,\ldots,5}=1$;
(b) (100) face of $C_{4v}$ NCS with $r=0.25$; (c) (100) face of $O$
NCS with $r=0.1$ and $g_2 = -1.5$; (d) (110) face of $T_{d}$ NCS with
$r=0.2$. The energy of the bound states is indicated by the color
scale, where black corresponds to zero energy and yellow to the
maximum energy $E_{\mathrm{max}}$; light grey indicates the absence of
a surface state. The light green line shows the projection of the
Fermi surface. As discussed in the text, in each case the zero-energy
state at ${\bf k}_\parallel=0$ is a topological helical Majorana. The
$C_2$ and $C_{4v}$ NCS also display topologically-protected arc states.}   
\end{figure}
%%%%%%%%%%%%%%%%%%%%%%%%%%%%

A similar (but more involved)
argument to that for the Weyl superconductors above
gives the implicit equation
~\cite{Schnyder12,Brydon11,EschrigIniotakisTanakabook}  for the bound states energies
\beqarray
0 & =& (\gamma^{+}_{{\bf p}^\prime} - \gamma^{-}_{{\bf p}})(\gamma^{-}_{{\bf
    p}^\prime} - \gamma^{+}_{{\bf p}})(|{\bf l}_{\bf p}||{\bf l}_{{\bf
    p}'}| - {\bf l}_{\bf p}\cdot{\bf l}_{{\bf p}'}) \nonumber \\
&& + (\gamma^{+}_{{\bf p}^\prime} - \gamma^{+}_{{\bf p}})(\gamma^{-}_{{\bf
    p}^\prime} - \gamma^{-}_{{\bf p}})(|{\bf l}_{\bf p}||{\bf l}_{{\bf
    p}'}| + {\bf l}_{\bf p}\cdot{\bf l}_{{\bf p}'}) .
\label{eq:cond2prop}
\eeqarray
where 
\beq
\gamma^{\pm}_{\bf k}
=
\frac{1}{\Delta^\pm_{\bf k}}\left[E -
  i \sgn(v_{F,\perp}({\bf k}))\sqrt{|\Delta^\pm_{\bf k}|^2 -
    E^2}\right]  \label{eq:gammapm}
\eeq
where $v_{F,\perp}({\bf k})$ is the component of the Fermi velocity in
the direction normal to the boundary.
Bound-state energies must lie within the minimum gap,
i.e. $|E|<\min\{|\Delta^{\pm}_{{\bf p}}|,|\Delta^{\pm}_{{\bf
    p}'}|\}$. Although the general solution must be obtained
numerically, analytic conditions can be derived 
for the existence of zero-energy states. Substituting
the zero-energy form of $\gamma^{\pm}_{{\bf k}} =
-i\sgn(v_{F,\perp}({\bf k})\Delta^{\pm}_{\bf k})$ 
into~\eq{eq:cond2prop} we obtain five conditions for zero-energy
states in terms of the spin-orbit coupling vector and the gap signs~\cite{Schnyder12}:
\begin{enumerate}
\item The spin-orbit vectors ${\bf l}_{{\bf p}}$ and ${\bf l}_{{\bf
    p}'}$ are parallel, and 
  the gap signs satisfy $\sgn(\Delta^{+}_{\bf p}) = \sgn(\Delta^{-}_{{\bf
      p}^\prime}) \neq \sgn(\Delta^{+}_{{\bf p}^\prime}) = \sgn(\Delta^{-}_{{\bf
      p}})$. 
\item The spin-orbit vectors ${\bf l}_{{\bf p}}$ and ${\bf l}_{{\bf
    p}'}$ are antiparallel, 
  and the gap signs satisfy $\sgn(\Delta^{+}_{\bf
    p}) = \sgn(\Delta^{+}_{{\bf p}^\prime}) \neq \sgn(\Delta^{-}_{{\bf
      p}^\prime}) = \sgn(\Delta^{-}_{{\bf 
      p}})$. 
\item The gap signs satisfy  $\sgn(\Delta^{+}_{\bf
    p}) = \sgn(\Delta^{+}_{{\bf p}^\prime})$ and $\sgn(\Delta^{-}_{{\bf
      p}^\prime}) \neq \sgn(\Delta^{-}_{{\bf p}})$. 
\item The gap signs satisfy  $\sgn(\Delta^{+}_{\bf
    p}) \neq \sgn(\Delta^{+}_{{\bf p}^\prime})$ and $\sgn(\Delta^{-}_{{\bf
      p}^\prime}) = \sgn(\Delta^{-}_{{\bf p}})$. 
\item The gap signs satisfy  $\sgn(\Delta^{+}_{\bf
    p}) \neq \sgn(\Delta^{+}_{{\bf p}^\prime})$, $\sgn(\Delta^{-}_{{\bf
      p}^\prime}) \neq \sgn(\Delta^{-}_{{\bf p}})$, $\sgn(\Delta^{-}_{{\bf
      p}^\prime}) \neq \sgn(\Delta^{+}_{{\bf p}})$, and $\sgn(\Delta^{+}_{{\bf
      p}^\prime}) \neq \sgn(\Delta^{-}_{{\bf p}})$. 
\end{enumerate}
These five conditions for a zero-energy state  include
the topological criteria defined  in Sec.~\ref{SecExampNCS}. 
Only a  state satisfying the first condition is not
topological: It
cannot be a Kramers-degenerate Majorana mode as the 
antisymmetric spin-orbit vectors at ${\bf p}$ and ${\bf p}'$ are not
parallel if ${\bf
  k}_{\parallel}$ lies at a time-reversal invariant point in the
surface Brillouin zone; Nor can it be an arc surface state protected by a
$\mathbb{Z}_2$ number, since the condition on the gap signs requires
that there be a node in the plane defined by ${\bf p}$ and ${\bf
  p}'$. On the other hand, since condition 
(ii) requires different sign of the positive and negative helicity
gaps, it is clearly consistent with the definition of the topological
  number~\eq{Z2number} for both these topological states. It also
describes nontopological 
states, however, as these conditions are not sufficiently restrictive.
The conditions (iii)-(v) exclusively describe flat band states
with nontrivial winding number, and they are equivalent to
the topological criteria~\eq{windingNoSurf}. Specifically, conditions (iii) and
(iv), which correspond to a situation where only one of the helical
gaps has a sign change between the two sides of the Fermi surface, describe 
topologically protected nondegenerate zero-energy states
with winding number $W_{(lmn)} = \pm1$. In contrast, (v)
requires sign changes of both helical gaps and gives doubly degenerate
states of winding number $W_{(lmn)}=\pm2$.

\subsubsection{Helical Majorana}

Helical Majorana states appearing at the centre of the surface
Brillouin zone (${\bf k}_\parallel = 0$) are relatively common for
majority-triplet NCS at high-symmetry surfaces. In each of the
examples shown in~\fig{dispersing} the zero-energy state at the zone
centre has this topological protection. The
other zero-energy states are present elsewhere in the surface
Brillouin zone in each of these examples, on the other hand, do not
satisfy the necessary criterion.

\subsubsection{Arc states}

Topologically protected arc states are rare, as the condition that
${\bf l}_{\bf k}$ be symmetric upon reflection about some plane places
severe constraints on the point group symmetry. There are two example
in~\fig{dispersing}: portions of the arc states at the $(110)$ surface
of the $C_{2v}$ NCS shown in panel (a), and the entire line of zero
energy states connecting the projection of the nodal rings at the
$(100)$ surface of the $C_{4v}$ NCS. In contrast, the arc states at
the $(110)$  surface of the $T_d$ NCS [panel (d)] do not have
topological protection, as the topological condition only holds for
planes which intersect gap nodes. 

%%%%%%%%%%%%%%%%%%%%%%%%%%%%
\begin{figure}
\begin{center}
\includegraphics[clip,width=\columnwidth]{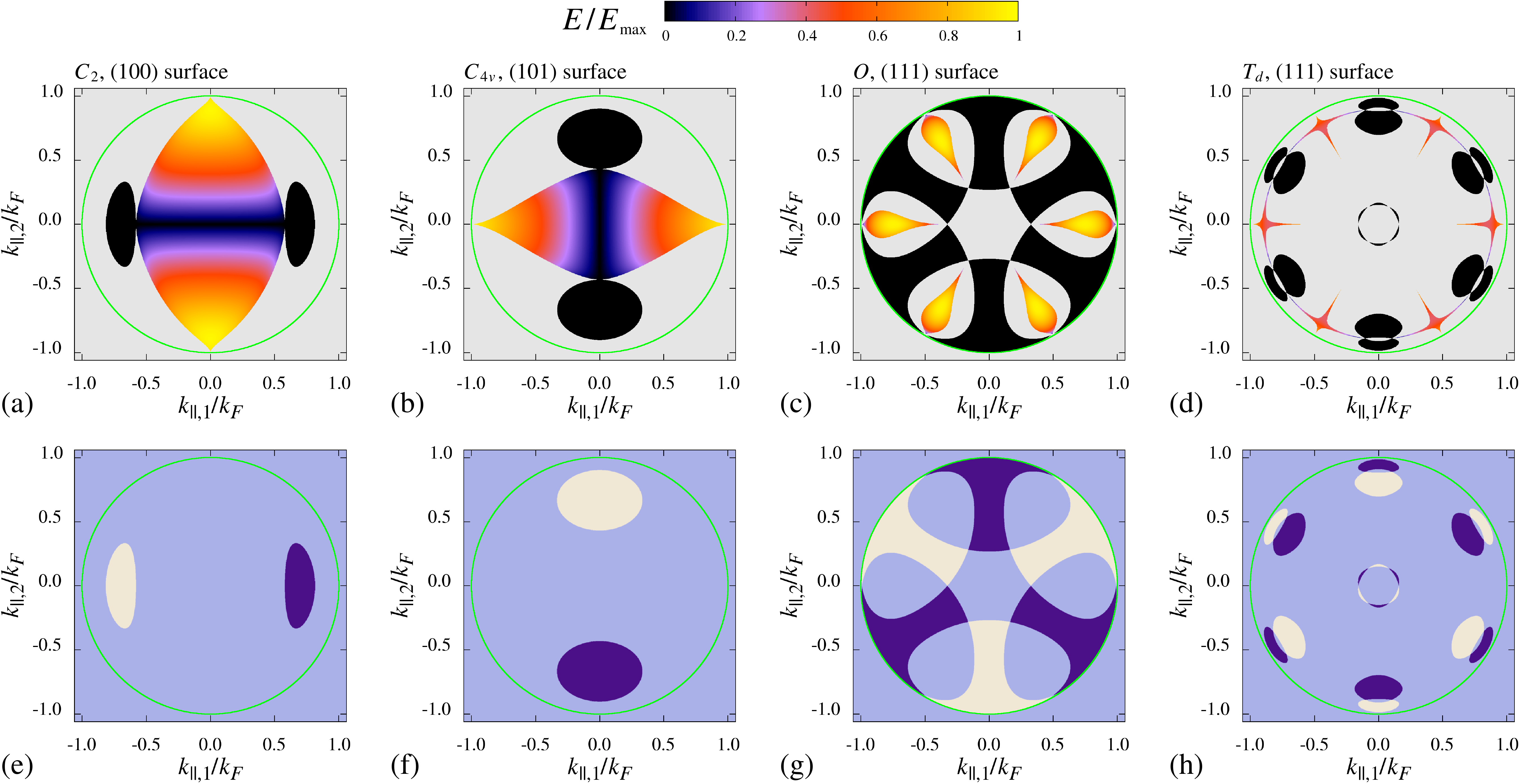}
\end{center}
  \caption{\label{zeroenergy} Topological flat-band states in nodal
    NCS. In the top row we plot the surface bound-state spectrum as a 
    function of surface momentum ${\bf k}_\parallel$ for (a) the
    (100) surface of a $C_{2}$ NCS with $r=0.25$ and
    $a_{i=1,\ldots,5}=1$, (b) the (101) face of a $C_{4v}$ NCS with
    $r=0.25$, (c) the (111) face of $O$ NCS with $r=0.25$ and $g_2 =
    -1.5$, (d) and the (111) face of $T_{d}$ NCS with $r=0.15$. The
    colour scale is 
    the same as in~\fig{dispersing}. In the bottom row we plot the
    corresponding variation of the 
    $W_{(lmn)}({\bf k}_\parallel)$.  Dark blue (gray) 
    corresponds to $W_{(lmn)}({\bf k}_\parallel)=+1$ ($-1$), while
    light blue represents. $W_{(lmn)}({\bf k}_\parallel)=0$. Figure
    (a), (b), (c) adapted from~\cite{brydonNJP2015}.}     
\end{figure}
%%%%%%%%%%%%%%%%%%%%%%%%%%%%

\subsubsection{Flat-band states}
\label{secFlatBandQuasiClassics}

Flat-band states are the most generic topological surface state of
nodal NCSs, appearing at any surface where the projection of the nodal
rings encloses a finite area and do not exactly lie on top of
another. We illustrate this with an example from each point group
in~\fig{zeroenergy}. In the top row we plot the spectrum in
the surface Brillouin zone,
while in the bottom row we show the corresponding variation of the
winding number $W_{(lmn)}({\bf k}_\parallel)$. The regions where the
topological number evaluates  to $\pm 1$ are bounded by the projected
line nodes of the gap, and exactly matches the location of the
zero-energy flat bands in the surface spectrum. By the bulk-boundary
correspondence this ensures the 
topological protection of the zero-energy flat bands and implies that
these states are nondegenerate.

%%%%%%%%%%%%%%%%%%%%%%%%%%%%
\begin{figure}
\begin{center}
\includegraphics[clip,width=0.6\columnwidth]{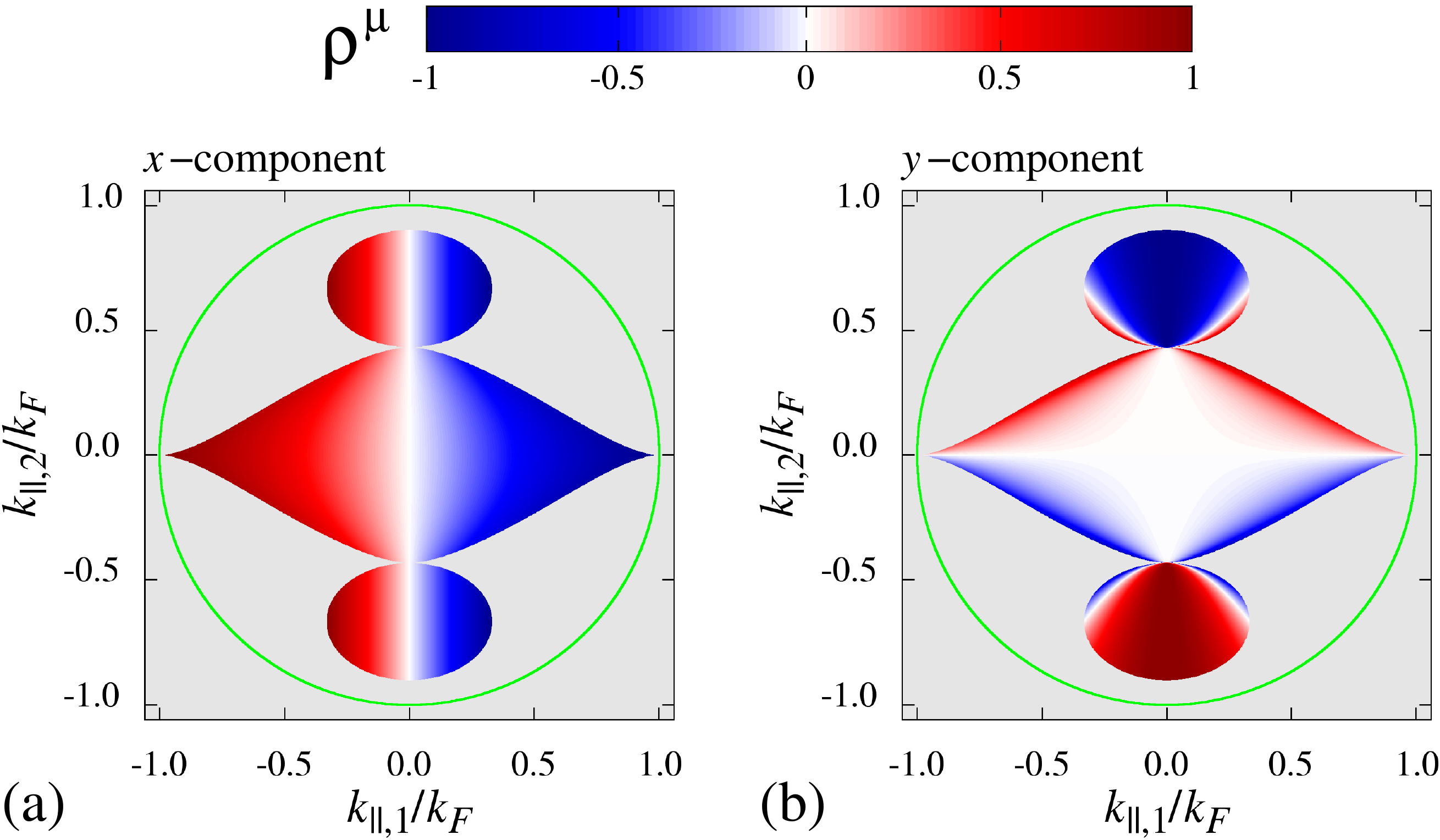}
\end{center}
  \caption{\label{spinPolarizationC4v} 
Spin polarization of the surface states at the (101) surface of a
$C_{4v}$ point group NCS with $r=0.25$. Panels (a) and (b) show the
spin polarization along the $x$- and $y$- axes of the crystal,
respectively. The $z$-spin polarization is vanishing. The energy of
the states is shown in~\fig{zeroenergy}(b).  Figure adapted
from~\cite{brydonNJP2015}. }   
\end{figure}
%%%%%%%%%%%%%%%%%%%%%%%%%%%%

\subsubsection{Spin polarisation of surface states}

The surface bound states of NCSs typically display a strong spin
polarization, which is a consequence of 
the broken inversion symmetry. The $\mu$-component of the spin
polarization of a bound state at transverse momentum ${\bf
  k}_\parallel$ and energy $E$ is defined~\cite{brydonNJP2015} 
as the expectation value 
\beq
\rho^{\mu}( E, {\bf k}_{\parallel}) = \int_{0}^{\infty} dr_\perp \Psi^{\dagger}({\bf
  k}_\parallel;r_\perp) \left(\begin{array}{c c}
\hat{\sigma}^{\mu} & 0 \cr
0 & - \left[ \hat{\sigma}^{\mu} \right]^{\ast}  \cr
\end{array}\right) \Psi({\bf
  k}_\parallel;r_\perp) \label{eq:spinpol}
\eeq
where $\Psi({\bf
  k}_\parallel;r_\perp)$ is the wavefunction.
Due to the symmetries of the Hamiltonian (time-reversal,
particle-hole, and chiral), the polarization of the edge states is
even in energy and odd in the surface momentum, i.e.
$\rho^{\mu}(E,{\bf k}_{\parallel}) = \rho^{\mu}(-E, {\bf
  k}_{\parallel}) = -\rho^{\mu}( E,-{\bf k}_{\parallel})$,
and there is no spin accumulation at the surface. The
spin polarization is sometimes defined only with reference to the
electron-like components of the  
wavefunction~\cite{tanakaNagaosaPRB09,vorontsovPRL08,luYip10}, but the
definition~\eq{eq:spinpol} is more reasonable as this quantity couples
to an applied Zeeman field.

The spin polarization of the edge states at the 
(101) surfaces of a $C_{4v}$ NCS is shown
in~\fig{spinPolarizationC4v}. Both the dispersing and the flat-band
edge states display a pronounced spin texture. The spin polarization
does not follow from the topology, however, but rather arises from the
unequal contribution of positive- and negative-helicity states to the
wavefunction. The spin polarization of the surface
states must match that of the continuum states where the two
intersect, in particular at the bounding nodes of the flat band states.
Despite its nontopological origin, the
spin polarization of the topological flat-band states is exploited in
several proposed experimental tests of their existence.

\section{Experimental signatures of topological surface states}\label{sec:signatures}

Directly detecting the surface states of a topological
  superconductor requires 
measurement techniques sensitive to the surface properties. This is a
much greater experimental challenge than  
demonstrating the existence of gap nodes in the bulk, as it generally
requires the preparation of high-quality surfaces or the incorporation
of the topological superconductor into a heterostructure
device. These hurdles are not insurmountable, however, as evidenced by
the observation of surface states in unconventional superconductors
such as the cuprates~\cite{KashiwayaPRB95,alffPRB97,weiPRL98} and 
Sr$_2$RuO$_4$~\cite{KashiwayaPRL2011}. In this section we examine
different experimental signatures of topological surface states.

\subsection{Tunneling conductance} 

Conductance spectroscopy experiments represent the dominant approach to
probing the surface states of unconventional
superconductors~\cite{Kashiwaya2000,Deutscher2005}. Since the conductance
is sensitive to the surface density of states of the
superconductor, it can provide direct evidence of surface states.
A number of authors have studied the conductance signatures of
topologically protected surface states of
NCS~\cite{Schnyder12,Brydon11,yadaPRB2011,tanakaPRL2010,EschrigIniotakisTanakabook,SchnyderPRB10,iniotakisPRB07,yamakage12,yokoyamaPRB2005,YuanPhysE2014,MukherjeeEPJB2011,brydonNJP2015,dahlhaus2012}.

The simplest model of a conductance experiment consists of a junction
between a topological superconductor and a normal metal with an
insulating barrier at their interface. The conductance
$\sigma_{S}(E)$  at zero
temperature is given by a generalization of the familiar
Blonder-Tinkham-Klapwijk formula~\cite{yokoyamaPRB2005,BTK}, 
\beq
\sigma_{S}(E) = \sum_{{\bf k}_\parallel}\left\{1 +
\frac{1}{2}\sum_{\sigma,\sigma'}\left(|a^{\sigma,\sigma'}_{{\bf
    k}_\parallel}|^2 - |b^{\sigma,\sigma'}_{{\bf k}_\parallel}|^2\right)\right\},
\eeq
where $a^{\sigma,\sigma'}_{{\bf k}_\parallel}$ and
$b^{\sigma,\sigma'}_{{\bf k}_\parallel}$ are the Andreev and normal
reflection coefficients, respectively, for spin-$\sigma$ electrons
injected into the superconductor at interface momentum ${\bf
  k}_\parallel$.  In the presence of spin-orbit coupling, both
spin-preserving ($\sigma'=\sigma$) and spin-flip ($\sigma'=-\sigma$)
Andreev and normal reflection coefficients are included. 
The scattering coefficients are determined by solution of the BdG
equation for an electron at bias energy $E=eV$ injected into the
superconductor from the normal lead. 
The insulating tunneling barrier is commonly modeled as a
$\delta$-function with strength $U =
ZE_F/k_F$~\cite{Kashiwaya2000}. The value of the 
dimensionless barrier parameter $Z$ defines different experimental
regimes: small values of $Z<1$ are least sensitive to surface states,
and are typically used to model 
point contact Andreev spectroscopy experiments~\cite{DagheroSST2010};
larger values $Z\gtrsim 1$ give clear signatures of the surface states
and are characteristic of tunnel junctions and scanning tunneling
microscopy. We will concentrate upon the latter regime.

%%%%%%%%%%%%%%%%%%%%%%%%%%%%
\begin{figure}
\begin{center}
\includegraphics[clip,width=0.7\columnwidth]{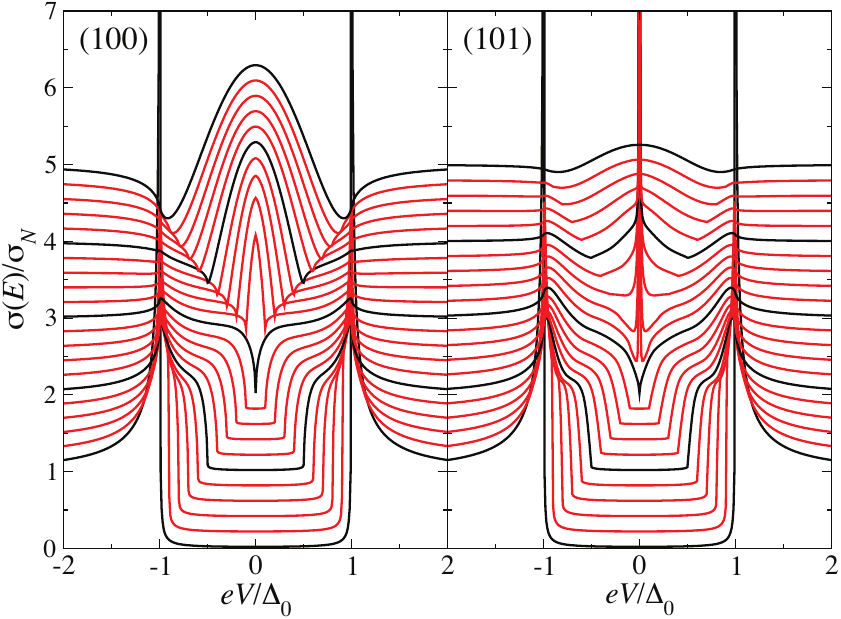}
\end{center}
  \caption{\label{tunnelConductanceC4v} 
Tunneling-conductance spectra at the (100) and (101) interfaces of a
$C_{4v}$ NCS with tunneling barrier strength $Z=4$. The results are
normalized by the normal state conductance $\sigma_N$ in the absence
of superconductivity. In both panels the singlet-triplet parameter $r$ is 
tuned from $r=0$ (purely triplet) at the top to $r=1$ (purely singlet)
at the bottom in steps of $0.05$. The curves at $r=0$, $0.25$, $0.5$,
$0.75$, and $1$ are highlighted in black. The conductance spectra are
offset by $0.2$ for clarity. }   
\end{figure}
%%%%%%%%%%%%%%%%%%%%%%%%%%%%

Tunneling conductance spectra at the $(100)$ and $(101)$
interfaces of a 
$C_{4v}$ NCS are shown in~\fig{tunnelConductanceC4v}. While the
majority-singlet cases ($r>0.5$) show a clear gap feature, 
majority-triplet pairing ($r<0.5$) is characterized by substantial
subgap conductance 
due to resonant tunneling through the surface states. At the $(100)$
interface, the dispersing arc states give a broad feature centred at
zero bias which completely fills the gap. This is also present at the
$(101)$ interface, but here the most striking aspect 
is the sharp zero-bias conductance peak due to tunneling into the
zero-energy flat-band states~\cite{Kashiwaya2000}. The sharp
zero-bias peak is a robust indicator of the flat-bands, as it arises  from
their singular contribution to the surface density of states. On the
other hand, fine structures in the dispersing states gives rise to
a much richer variation in their contribution to the conductance, and
there is no universal signature~\cite{Brydon11,EschrigIniotakisTanakabook}.
In particular, there is no way to distinguish the dispersing
surface states of a NCS from those of a Weyl superconductor using
simple tunneling conductance measurements.

Many variations on the basic junction geometry are possible,
e.g., replacing the normal lead by a metallic
ferromagnet~\cite{Annunziata2012,MukherjeeEPJB2011,WuPRB2009}, or  
assuming a spin-active barrier~\cite{brydonNJP2015}. This allows
conductance measurements to probe the spin structure and, indirectly,
the degeneracy of the surface states. Applying a Zeeman field to
the sample is predicted to shift the spin-polarized flat bands away
from zero energy, thus splitting the zero-bias
peak~\cite{yadaPRB2011,matsuura2012,wong13}. The absence of 
splitting for certain field orientations would strongly suggest
nondegenerate bands, as doubly-degenerate bands are expected
to be spin degenerate and split for any orientation of the applied
field~\cite{KashiwayaPRB1999}. On the other hand, the stability of
nodal NCSs in such Zeeman fields is
uncertain~\cite{EreminPRB2006,LoderJPCM2013}. A less invasive  
variation on this approach is to replace the insulating tunneling
barrier with a ferromagnetic
insulator~\cite{KashiwayaPRB1999,brydonNJP2015}. 

\begin{figure}
\begin{center}
\includegraphics[clip,width=0.7\columnwidth]{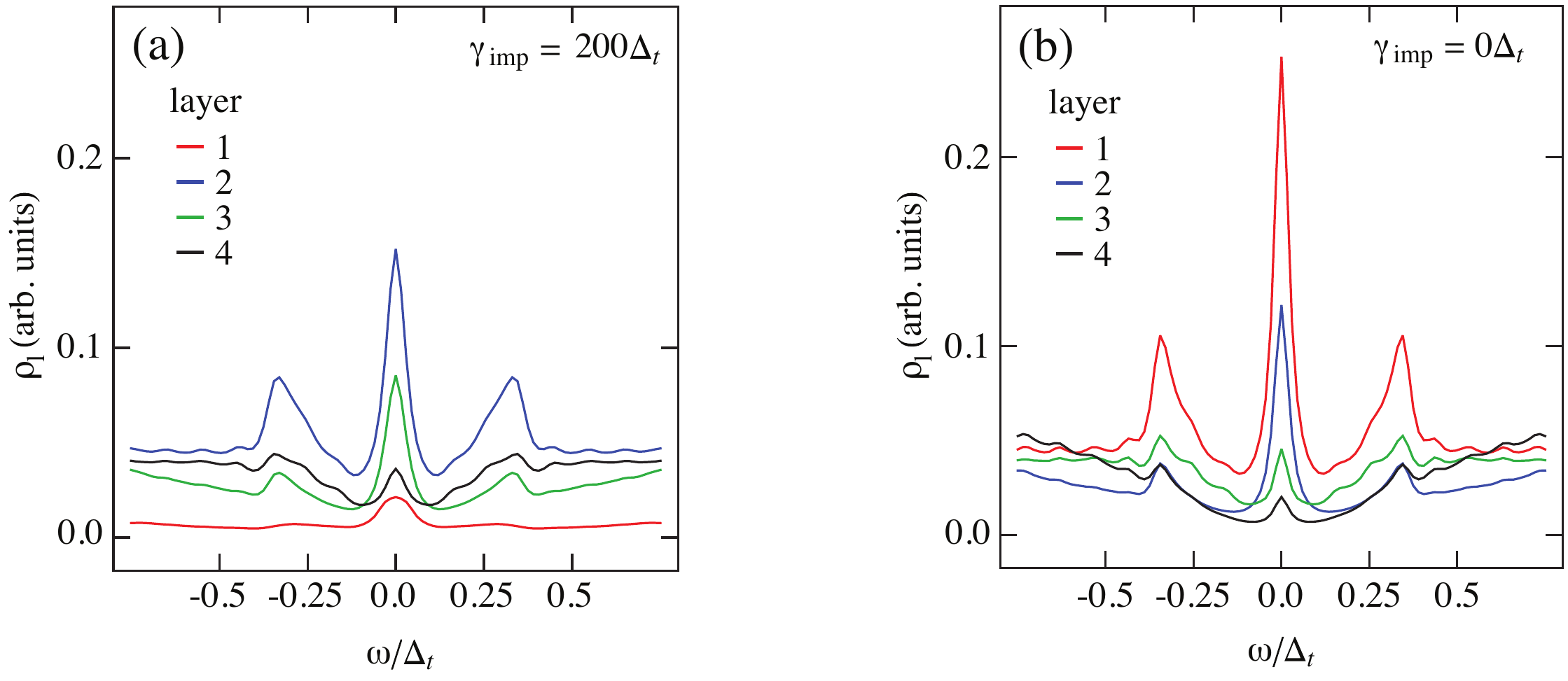}
\end{center}
\caption{\label{disorderFig} 
Layer-resolved local density of states $\rho_l$ of a
$(d_{xy}+p)$-wave NCS in a ribbon geometry with (01) edges, which
displays zero-energy flat bands.
Panel (a) corresponds to a dirty 
NCS with non-magnetic Gaussian disorder in the outermost layer of strength $\gamma_{\mathrm{imp}} = 200 \Delta_t$,
while panel (b) represents a clean NCS. In the former, the outermost
layer is ``dead'' and the topologically protected edge states develop
in the second layer.
Figure adapted from Ref.~\cite{QueirozPRB2014}.}   
\end{figure}

The topological classification of nodal superconductors is only
approximately valid in the presence of disorder. Even for a pristine
bulk system, however, disorder at the surface or at the interface in 
a tunneling junction is inevitable. So long as the disorder is not
too strong the conductance features discussed above remain
detectable, although they are broadened by the finite lifetime of the
quasiparticles~\cite{MatsumotoJPSJ1995,BarashPRB1997,dahlhaus2012,QueirozPRB2014,QueirozPRB2015}. Extremely    
disordered surfaces produce ``dead'' insulating layers, which
redefine the edge of a 
superconductor~\cite{QueirozPRB2014,QueirozPRB2015,SchubertPRB2012}.
This is illustrated in Figs.~\ref{disorderFig}(a)
and~\ref{disorderFig}(b), which show the layer-resolved density of
states of a $(d_{xy}+p)$-wave NCS~\cite{tanakaPRL2010} on a
  square lattice in the presence 
and absence of strong edge disorder, respectively. For very strong
edge impurities, the states in the outermost layer are almost fully
localized, while in the second and third inward layers new weakly
disordered states reemerge.

%%%%%%%%%%%%%%%%%%%%%%%%%%%%
%%%%%%%%%%%%%%%%%%%%%%%%%%%%

\subsection{Surface currents} 
\label{SecSurfaceCurrents}

A key signature of the chiral surface states of Weyl
superconductors is that they show anomalous spin and thermal Hall
conductance in the direction of their
propagation~\cite{meng12,VolovikJPCM1989}. Although the
zero-temperature Hall
conductances are directly related to the Chern number, they also depend
upon the number of chiral edge modes in the system, and so they are
proportional to the separation 
of the Weyl points in momentum space. This is in contrast to the universal
quantization in a fully-gapped time-reversal-symmetry breaking
topological
superconductor~\cite{SenthilPRB1999,ReadGreenPRB2000,SenguptaPRB2006}. Furthermore, 
the spin Hall conductance in    
the triplet Weyl superconductor depends on the orientation of the
magnetic field with respect to the ${\bf d}$-vector, as not all spin
directions are conserved~\cite{SenguptaPRB2006}.
While the measurement of the anomalous spin Hall conductance is likely
to be very difficult~\cite{SenthilPRB1999}, the anomalous thermal Hall
conductance is predicted to be experimentally accessible for realistic
materials~\cite{goswamiArXiv13,goswamiArXiv14}.

The chiral surface states may also
carry a surface charge current, but it is not topologically
protected since charge is not conserved in a superconductor. In
particular, its detection is 
extremely difficult as it will be compensated by Meissner screening
currents~\cite{MatsumotoJPSJ1999}. While the 
current carried by the bound states is localized within a few
coherence lengths $\xi$ of the surface, however, the screening
currents build up on the scale of the penetration depth
$\lambda$, and so in extreme type-II superconductors a finite
surface current density is expected. Precision
magnetometry has so far failed to detect surface charge currents in
Sr$_2$RuO$_4$~\cite{KirtleyPRB2007}, however, despite otherwise compelling
evidence that this is a chiral $p$-wave
superconductor~\cite{maenoJPSJ12,mackenzieRMP03}. The charge current
is not universally related to the Chern number, however, and even
ignoring the Meissner effect, it
depends sensitively upon microscopic details of the
superconductor~\cite{LedererPRB2014,HuangPRB2014}.

Surface currents might also arise in time-reversal
invariant NCS. Ignoring spin-orbit coupling, the simplest model of a
majority-triplet $C_{4v}$ NCS is a superconducting analogue of
the quantum spin Hall insulator~\cite{satoPRB06}, and the helical edge
states therefore 
carry a surface spin current~\cite{vorontsovPRL08,SatoFujimotoPRB2009,tanakaNagaosaPRB09}. 
On the other hand, a surface spin current carried by states above the gap
is present in the nontopological phase~\cite{luYip10}, and the spin
current is strongly suppressed by the 
spin-orbit coupling~\cite{vorontsovPRL08}. Charge currents are also
predicted to arise at the interface between NCSs and
ferromagnets~\cite{brydonNJP2013,Schnyder13,RenEPJB2013}. In particular, at 
surfaces with spin-polarized flat bands, the coupling to the
exchange field of the ferromagnet generates a chiral
electronic structure at the interface. The resulting charge current
is very pronounced at low  
temperatures and shows nonanalytic dependence on the exchange field
strength~\cite{brydonNJP2013,Schnyder13}.

\subsection{Other proposals}

There are a number of proposals for evidencing chiral or
helical edge states based upon interference effects in the
magnetoconductance of superconductor
loops~\cite{BeriPRB2012,DiezPRB2013}, or nonlocal conductance at
domain walls in Weyl superconductors~\cite{SerbanPRL2010}. Although
such experiments could give unambiguous demonstration of the edge
states, these proposals assume a fully-gapped pairing state, so it is 
unclear if they would apply to a nodal system.

\begin{figure}
\begin{center}
\includegraphics[clip,width=0.6\columnwidth]{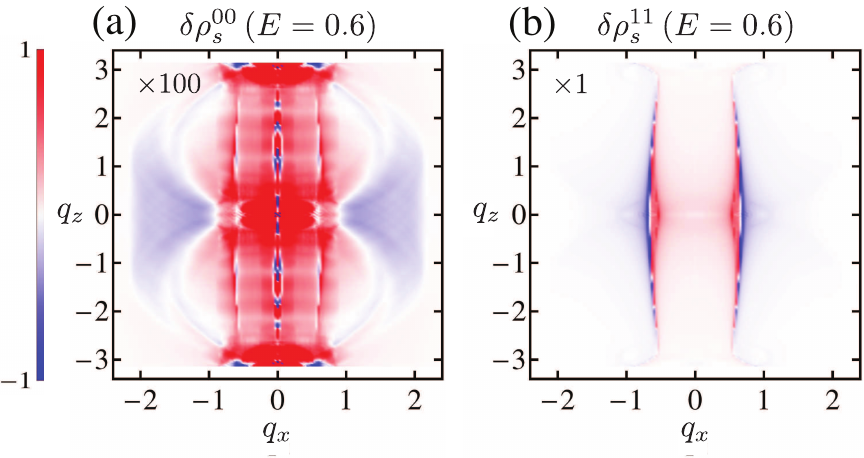}
\end{center}
  \caption{\label{QPIpatternC4v} Fourier-transformed
    quasiparticle interference patterns for arc states at the
    (010) surface of a
    $C_{4v}$ point group NCS. Results for (a) nonmagnetic scatterers
    and (b) magnetic scatterers with impurity moment along the
    $x$-axis. The weak and diffuse features in (a) reflect the
      suppression of backscattering due to the
      spin-polarization of the edge states; backscattering occurs for
      magnetic scattering, giving rise to the much stronger and
      sharper features in (b). Figure adapted from
      Ref.~\cite{hofmannPRB13}.} 
\end{figure}

The surface states of a topological
superconductor could be more directly demonstrated using
quasiparticle interference 
spectroscopy~\cite{hofmannPRB13,SauTewariFMSC2012}, which has already been
successfully used to evidence topological insulator 
surface states~\cite{RoushanNature2009}. This can probe the
spin character of the surface states, as scattering between states
with opposite spin polarization (i.e. backscattering processes) are 
suppressed, unless the scattering is by magnetic
impurities. Predictions~\cite{hofmannPRB13} for
NCS arc states are shown in Fig.~\ref{QPIpatternC4v}: whereas
nonmagnetic scattering gives a weak and diffuse signal [panel
    (a)], magnetic scattering produces a strong and well-defined
image of the arc states [panel (b)].

%%%%%%%%%%%%%%%%%%%%%%%%%%%%
%%%%%%%%%%%%%%%%%%%%%%%%%%%%
   
\section{Survey of candidate materials for nodal topological superconductivity}
\label{sec:materialsurvey}

In certain materials with strong electronic correlations, 
topological superconductivity can occur naturally
due to unconventional~\cite{normanReivewArxiv14},
i.e.~non-phonon-mediated, pairing interactions. 
These unconventional pairing mechanisms (e.g., due to spin
fluctuations) are in general quite anisotropic and hence give 
rise to nodal structures of the superconducting gap. As a consequence,
topological superconductors with nodes are far more common than those
with a full excitation gap.
An important exception to this rule is the fully gapped spin-triplet
superconductor Sr$_2$RuO$_4$~\cite{maenoJPSJ12,mackenzieRMP03}.
In this section we give a brief overview of candidate materials
  for nodal topological 
superconductivity with either spin-triplet or spin-singlet pairing,
which we list in Table~\ref{TabCandidates} 
and Table~\ref{TabCandidatesSinglet}, respectively. We note that due
to strong atomic spin-orbit coupling in many of these materials, 
the spin symmetry of the Cooper pairs is determined by the
\emph{pseudospins} of  quasiparticles 
with opposite momenta, rather than by the real spins of electrons. 
Thus, the pairing symmetries may be better distinguished in terms of
the parity of the orbital pair wave function; for simplicity, we 
nevertheless refer to odd- and even-parity states as spin triplet and
singlet, respectively.

\subsection{Nodal superconductors with spin-triplet pairing}

A nodal spin-triplet pairing state has been established unambiguously
in superfluid ${}^3$He~\cite{wheatleyRMP,vollhardtBook}.  
That is, the A phase of ${}^3$He is known to be  described by the
Anderson-Brinkman-Morel (ABM) state~\cite{andersonMorel}, 
a three-dimensional superfluid  with nodal points at the north and
south poles of the Fermi sphere.  
The ABM state is a member of  symmetry class D with codimension $p=3$
(cf.~Sec.~\ref{SecExampChiralP}).  
Hence, according to the classification of
Table~\ref{classificationTable},  the point nodes of ${}^3$He-A are
topologically stable and protected by a two-dimensional Chern
number~\cite{Volovik:book,VolovikExotic,volovikLectNotes13,Volovik3HeA}. 

Here, however we want to focus on nodal topological
\emph{superconductors}  rather than superfluids. 
A very promising group of materials for nodal topological
superconductivity are the noncentrosymmetric superconductors
(NCSs)~\cite{bauerSigristBook,yipReview14}. 
These systems have extremely diverse properties, however, as
they are defined by their crystal structure as opposed to chemical
composition. In particular, the parity-mixing of the gap depends upon
microscopic details. The most interesting case for us is 
when the spin-triplet and spin-singlet components are comparable in
magnitude, which generically leads to a nodal superconducting state as
discussed in Sec.~\ref{topoSection}.
Sizable spin-triplet pairing components can be expected to occur in
strongly correlated NCSs where the onsite Coulomb repulsion suppresses
the spin-singlet $s$-wave channel, and where there are strong
noncentrosymmetric spin
fluctuations~\cite{YokoyamaPRB2007,yanaseSigristJPSJ08,TakimotoJPSJ2009}. 
These conditions are believed to hold in the heavy fermion $C_{4v}$
point group NCS
CePt$_3$Si~\cite{bauerPRL04}, leading to an $(s+p)$-wave
state~\cite{yanaseSigristJPSJ08}. Indeed, there is significant
evidence of line nodes and an unconventional spin pairing state in this
material~\cite{bonaldePRL05,IzawaPRL05,mukudaJPSJ09,takeuchiJPSJ07}.
A similar pairing state seems to be realized in
CeIrSi$_3$~\cite{mukudaPRL08,MukudaPRL10} and 
CeRhSi$_3$~\cite{kimuraPRL07}, which are also heavy fermion NCS with
$C_{4v}$ point group.
Not all NCS with sizable spin-triplet gap are so strongly
correlated: Replacing Pd by Pt in the $O$ point group NCS
Li$_2$(Pd$_{1-x}$Pt$_x$)B 
appears to tune the system from a conventional $s$-wave superconductor
to a nodal triplet
superconductor~\cite{yuanPRL06,nishiyama07,EguchiPRB2013}, but there
is no evidence that the correlation strength is dramatically altered
by this substitution. The gap structure of LaNiC$_2$ ($C_{2v}$ point group) is
controversial~\cite{hillierPRL09,bonaldeLee2011,ChenNJP2013}. Muon
spin rotation experiments evidence time-reversal-symmetry
breaking in LaNiC$_2$~\cite{hillierPRL09} and the related
(centrosymmetric) LaNiGa$_2$~\cite{HillierPRL12}, suggesting
nonunitary triplet pairing.

\begin{table}[t!]
\centering
\caption{List of candidate materials for nodal topological
  superconductivity with (majority) spin-triplet pairing. Note that in some
  materials the evidence is contradictory.
NCS: Noncentrosymmetric superconductor. HF: Heavy fermion superconductor. FM: Ferromagnetic superconductor. 
NMR: Nuclear magnetic resonance.
SH: Specific heat. UA: Ultrasound attenuation. TC: Thermal
conductivity. PD: London penetration depth. $H_{c2}$: Upper
  critical field }
\label{TabCandidates}
\begin{center}
\begin{threeparttable}
\begin{tabular}{| c | c c c c  |}
\hline
Material & Type  &  $\begin{array}{c}
\mbox{Evidence for} \\
\mbox{triplet pairing}
\end{array}$ & Evidence for nodes &  
$\begin{array}{c}
\mbox{Probable pair-} \\
\mbox{ing symmetry}
\end{array}$  \cr
\hline  
 \hline
A phase of ${}^3$He & superfluid & NMR, magnetiz.~\cite{wheatleyRMP} & SH~\cite{vollhardtBook}  & chiral  \cr
\hline
CePt$_3$Si & NCS, HF & indirect & PD, NMR, etc.~\cite{bonaldePRL05,IzawaPRL05,mukudaJPSJ09,takeuchiJPSJ07}    & ($s+p$)-wave  \cr
CeIrSi$_3$${}^{\footnotesize{\textrm{$\dag$}}}$ & NCS, HF & NMR~\cite{mukudaPRL08,MukudaPRL10} & NMR~\cite{mukudaPRL08,MukudaPRL10}   & ($s+p$)-wave \cr  
CeRhSi$_3$${}^{\footnotesize{\textrm{$\dag$}}}$ & NCS, HF & $H_{c2}$~\cite{kimuraPRL07} &   & ?\cr  
Li$_2$Pt$_3$B & NCS & NMR~\cite{nishiyama07} & PD, NMR, SH~\cite{yuanPRL06,nishiyama07,EguchiPRB2013}  & ($s+p$)-wave\cr  
LaNiC$_2$ & NCS & indirect~\cite{hillierPRL09} &
PD~\cite{bonaldeLee2011}  &  nonunitary \cr
LaNiGa$_2$ & centro. & 
  indirect~\cite{HillierPRL12} &  &  nonunitary \cr
\hline 
URhGe& FM, HF & indirect~\cite{hardyPRL05} & SH~\cite{huxleyNature2001}   &  $p$-wave \cr
UCoGe & FM, HF & NMR~\cite{hattoriJPSJ14} & indirect~\cite{HuyPRL08}   &  $p$-wave \cr
UGe$_2$${}^{\footnotesize{\textrm{$\dag$}}}$ & FM, HF & $H_{c2}$~\cite{saxenaNature00,SheikinPRB2001} &  NMR~\cite{haradaPRB07}   & $p$-wave   \cr
\hline
UPt$_3$ & HF   & NMR \cite{touPRL98} & SH, UA, TC \cite{UPt3RMP}  &
chiral $f$-wave  \cr
UBe$_{13}$ & HF   & NMR \cite{Tou2007706} & SH, NMR~\cite{ottPRL84a,ottPRL84b}  &   nodal  \cr
\hline
\end{tabular}
\begin{tablenotes}
\item    \item[${}^{\textrm{$\dag$}}$]  superconducting under pressure
\end{tablenotes}
\end{threeparttable}
\end{center}
\end{table}

A number of Uranium-based heavy fermion compounds show strong evidence
of unconventional triplet pairing states with nontrivial topological
properties~\cite{flouquetPolonica03,Brison2000165,RiseboroughBook,pfleidererRMP09}. 
Among them, UPt$_3$ is probably the most promising candidate for
spin-triplet superconductivity, and has received much
attention~\cite{UPt3RMP}. This material has three  
different nodal superconducting phases, and there is strong evidence
that one of these realizes a time-reversal-symmetry breaking
chiral $f$-wave state, which is predicted to show Weyl
nodes~\cite{goswamiArXiv14,StrandPRL2009}.   
The superconductivity of UBe$_{13}$ is less well-understood, but
evidence of surface states has been reported~\cite{waettiOttPRL00}.
Intriguingly, some of these systems, in particular URhGe, UCoGe, and
UGe$_2$, show an apparently cooperative coexistence of ferromagnetism
and superconductivity~\cite{flouquetPolonica03,pfleidererRMP09}. That
is, close to a ferromagnetic quantum 
critical point ferromagnetic spin fluctuations mediate spin-triplet
pairing. The natural pairing state for such systems would involve
equal-spin-pairing   
 $\left| \uparrow \uparrow \right\rangle$  and $\left| \downarrow
\downarrow \right\rangle$ on the majority- and minority-spin Fermi
surfaces, respectively. The crystal space group of most ferromagnetic
superconductors is of low symmetry, which results in strong
anisotropies of the magnetic fluctuations, favourable to the observed
nodal pairing states. Topological properties of these systems are
largely uninvestigated, although the existence of arc states connected
with Weyl nodes has been proposed~\cite{SauTewariFMSC2012}.

The relative rarity of intrinsic spin-triplet superconductors has lead
to a number of proposals to engineer such states in
heterostructures involving spin-singlet
superconductors~\cite{meng12,das13,takeiPRB13,LinderPRL2010,fanMelePRL2013}.  
A key to these proposals is the presence of strong spin-orbit
coupling, which, under appropriate conditions, can accomplish a change
in parity of the Cooper pairs, resulting in a sizable
proximity-induced spin-triplet gap. For example, superlattices of
topological insulators and $s$-wave superconductors are proposed to
realize a $p$-wave Weyl
superconductor~\cite{meng12,das13}. Incorporating 
high-temperature superconductors into such devices has
the great advantage that the engineered spin-triplet superconductor
could be realized at much higher temperatures than their intrinsic
counterparts~\cite{takeiPRB13,LinderPRL2010,fanMelePRL2013}.
Particularly interesting is proximity-induced pairing in 
the surface states of a topological
insulator~\cite{FuKane_SC_STI,LinderPRL2010}. Recently, 
cuprate superconductor-topological insulator devices have been
fabricated~\cite{WangNatPhys2013}.  However, the proximity-induced gap
in these devices is found to be nodeless.

\subsection{Nodal superconductors with spin-singlet pairing}

High-temperature cuprate superconductors are probably the most
extensively studied unconventional
superconductors~\cite{normanReivewArxiv14}.  
While there is no common consensus on the pairing mechanism, it is by
now generally accepted that the pairing symmetry in the cuprates  is
of $d_{x^2-y^2}$-wave form~\cite{Scalapino1995329,tsueiRMP00}. That
is, the superconducting gap changes sign along the Fermi surface,
giving rise to four topologically protected point  
nodes. The superconducting state of the cuprates belongs to symmetry
class CII in Table~\ref{classificationTable}, since it preserves both
time-reversal and $SU(2)$ spin-rotation symmetry.
Hence, it follows from the classification
of Table~\ref{classificationTable}, that the point nodes  
are protected by a $2 \mathbb{Z}$ topological number, i.e., by a
winding number which takes on the values $\pm 2$.  
Due to the bulk-boundary correspondence, the topological
characteristics of these point nodes lead to the appearance of  
doubly degenerate (i.e., spin-degenerate) flat band edge
states~\cite{RyuHatsugaiPRL02,huPRL94}.   
These zero-energy edge modes have been observed in tunneling
experiments on the cuprate material YBa$_2$Cu$_3$O$_{6-x}$
\cite{KashiwayaPRB95,alffPRB97,weiPRL98}. 

\begin{table}[t!]
\centering
\caption{List of candidate materials for nodal topological
  superconductivity with (majority) spin-singlet pairing.
HF: Heavy fermion superconductor. SL: Superlattice.
ARPES: Angle-resolved photoemission spectroscopy.
STM: Scanning tunneling microscopy. NMR: Nuclear magnetic resonance. PD: London penetration depth.
SH: Specific heat. TC: Thermal conductivity.  MT: Magnetic
torque. PKE: Polar Kerr effect. $\mu$SR: Muon spin rotation.}
\label{TabCandidatesSinglet}
\begin{center}
\begin{threeparttable}
\begin{tabular}{| c | c c c c  |}
\hline
Material & Type  &  $\begin{array}{c}
\mbox{Time-reversal} \\
\mbox{symmetry}
\end{array}$ & Evidence for nodes &  
$\begin{array}{c}
\mbox{Probable pair-} \\
\mbox{ing symmetry}
\end{array}$  \cr
\hline  
 \hline
 $\begin{array}{c}
\mbox{YBa$_2$Cu$_3$O$_{6+x}$,}\\
 \mbox{La$_{2-x}$Sr$_x$CuO$_4$, etc.} 
 \end{array}$
  & 
   $\begin{array}{c}
\mbox{ high-temp.}\\
 \mbox{supercond.} 
 \end{array}$
 & Yes & 
$\begin{array}{c}
\mbox{ARPES, STM,} \\
\mbox{NMR, PD, etc.~\cite{Scalapino1995329,tsueiRMP00}} 
\end{array}$
 & $d_{x^2-y^2}$-wave  \cr
\hline
CeCu$_2$Si$_2$ & HF & Yes &  indirect~\cite{steglichPRL11}    & $d$-wave  \cr
CeCoIn$_5$ & HF & Yes  &   $\begin{array}{c}\mbox{SH, TC, NMR,} \\ \mbox{STM~\cite{IzawaPRL01,MovshovichPRL01,allanDavis13,zhouYazdani13,KohoriPRB01}}  \end{array}$    & $d_{x^2-y^2}$-wave  \cr
CeIrIn$_5$ & HF & Yes  &  SH, TC, NMR~\cite{MovshovichPRL01,KohoriPRB01}    & $d$-wave  \cr
CeRhIn$_5$${}^{\footnotesize{\textrm{$\dag$}}}$ & HF & Yes  &  SH~\cite{FisherPRB02}    & $d$-wave  \cr
URu$_2$Si$_2$ & HF &  
 $\begin{array}{c}
 \textrm{No (MT~\cite{liBalicas13}},
 \\
 \textrm{PKE~\cite{SchemmArxiv14})}
 \end{array}$
&  SH, NMR, TC~\cite{kasahara07,YanoMachidaPRL08}    & $(d \pm id)$-wave  \cr
\hline
SrPtAs & pnictide & No ($\mu$SR~\cite{biswasPRB2013}) &
indirect~\cite{fischerPRB14}    & ($d  \pm i d $)-wave  \cr 
\hline
CeCoIn$_5$/YbCoIn$_5$ & SL
& & indirect~\cite{ShimozawaPRL2014} &  likely $d$-wave \cr
Cu$_x$(PbSe)$_5$(Bi$_2$Se$_3$)$_6$ & SL & &
SH~\cite{SasakiPRB2014} & line node \cr
\hline
\end{tabular}
\begin{tablenotes}
\item    \item[${}^{\textrm{$\dag$}}$]  superconducting under pressure
\end{tablenotes}
\end{threeparttable}
\end{center}
\end{table}

A superconducting gap with $d$-wave symmetry also occurs in the
Cerium-based heavy fermion compounds CeCu$_2$Si$_2$~\cite{SteglichPRL79} 
and CeXIn$_5$,  where X is a transition metal (Co, Ir,
Rh)~\cite{IzawaPRL01,MovshovichPRL01}. 
In all these materials superconductivity appears close to an antiferromagnetic quantum critical point and   is believed to
be mediated by spin
fluctuations~\cite{RiseboroughBook,pfleidererRMP09,normanReivewArxiv14}. 
In contrast to the ferromagnetic superconductors, however, magnetic
order in the Cerium compounds competes with superconductivity.
NMR measurements indicate spin-singlet pairing and in combination with
specific heat and 
thermal conductivity measurements yield evidence for line nodes in the
superconducting gap. 
Therefore, it seems likely that the order parameter symmetry in these
heavy fermion superconductors is of $d$-wave form. 
This has been confirmed for CeCoIn$_5$ by a careful analysis of
the quasiparticle interference patterns 
observed in scanning tunneling spectroscopy~\cite{allanDavis13,zhouYazdani13}.

The table~\ref{TabCandidatesSinglet} lists two compounds which show
signs of unconventional spin-singlet superconductivity with broken
time-reversal symmetry.  The first one is
URu$_2$Si$_2$~\cite{mydoshRMP11}, in which superconductivity emerges
out of a somewhat mysterious ``hidden order phase". Recently, a polar
Kerr effect has been observed, which appears  with the onset of
superconductivity~\cite{SchemmArxiv14}. Together with specific
heat~\cite{YanoMachidaPRL08} and thermal conductivity
measurements~\cite{kasahara07}, this suggests that URu$_2$Si$_2$ is a
chiral $d$-wave superconductor, which breaks time-reversal
symmetry. Such a state can support Weyl
  nodes~\cite{goswamiArXiv13}. The second candidate for time-reversal
  symmetry breaking superconductivity is 
the pnictide material SrPtAs. Muon spin-rotation measurements on
SrPtAs have revealed time-reversal-symmetry breaking in the
superconducting state~\cite{biswasPRB2013}. This material is also
predicted to realize a chiral $d$-wave state with Weyl
nodes~\cite{fischerPRB14}. 
 
Finally, we comment on artificial superconducting
structures.  Recently, superlattices of the 
$d_{x^2-y^2}$-wave superconductor CeCoIn$_5$ and the
non-superconducting metal YbCoIn$_5$ have been grown, such that
inversion symmetry is broken in the CeCoIn$_5$
layers~\cite{ShimozawaPRL2014}. This material is predicted to realize
a nodal NCS with nondegenerate flat 
bands~\cite{tanakaPRL2010,YuanPhysE2014}. Another intriguing compound is
Cu$_x$(PbSe)$_5$(Bi$_2$Se$_3$)$_6$, where the doped topological insulator
Cu$_x$Bi$_2$Se$_3$ is combined with the topologically trivial insulator 
PbSe~\cite{SasakiPRB2014}. Cu$_x$Bi$_2$Se$_3$ has attracted much
attention as a candidate topological superconductor with a full
(odd-parity) gap~\cite{FuBergPRL2010,KrienerPRL2011,Levy2013}. Surprisingly,
specific heat measurements appear to indicate that the superconducting gap in
Cu$_x$(PbSe)$_5$(Bi$_2$Se$_3$)$_6$ has line nodes. Note that we have
conservatively classified the pairing as spin singlet, although the
upper critical field hints at a more exotic gap symmetry.

\section{Summary and outlook}

In this review we have presented an introduction to the topological
properties of nodal superconductors. Although the topology is less
robust than in fully-gapped superconductors, nodal superconductors are
more common than unconventional fully-gapped superconductors, and
so this theory will likely find much application.
In addressing this subject, we have introduced two important
examples: time-reversal-symmetry-breaking Weyl superconductors
(Sec.~\ref{sec:topSCwpointnodes}), and 
time-reversal-invariant noncentrosymmetric superconductors
(Sec.~\ref{sec:topSCwlinenodes}), which can 
be regarded as prototypes of superconductors with topologically
nontrivial point and line nodes, respectively. 

At the heart of the review is Sec.~\ref{topoSection}, and more
specifically the topological classification of nodal superconductors
summarized in Table~\ref{classificationTable}. We have organized
our review around this classification scheme, since: (i) It brings
order to the growing zoo of unconventional superconductors; 
(ii) it gives guidance for the search and design of new topological
states; and (iii) it links the properties of the surface states to the
bulk nodal structure and the symmetries of the order parameter. Point
(iii) is particularly exciting, as it means that the pairing symmetry of
an unconventional superconductor can be identified solely by means of
surface measurements. Accurate modelling of the surface state spectrum
is therefore an 
important aspect of the theory. This is discussed from the point of
view of the popular quasiclassical approximation in
Sec.~\ref{sec:quasiclassical}. While the theory agrees perfectly with
the topological arguments, it also gives us access to physics beyond
 Table~\ref{classificationTable}, e.g.,  edge states
without topological protection, and the physical attributes of all
types of surface states. The latter is very useful input for
designing experimental probes of the surface.

In Sec.~\ref{sec:materialsurvey} we briefly surveyed a number of
materials for which there is strong evidence of a nodal
superconducting state with nontrivial topology. Demonstrating
the existence of the topologically protected edge states, however,
will be a significant experimental challenge. As we have discussed in 
Sec.~\ref{sec:signatures}, there have been many theoretical
studies of the possible signatures of topological edge states, but
only a few materials have been practically studied. While some
systems are likely less suitable for such
studies than others due to material-specific difficulties, the large
number and diversity of candidate 
superconductors gives hope that direct detection of edge states is
feasible. In this regards, proposals to realize a topological nodal
superconducting state in heterostructures are very attractive, due to
the fine experimental control over such systems. We expect this
  to be a major direction for both theoretical and experimental work.
Another promising avenue for future research is to utilize 
the edge states of topological superconductors to
 design novel quantum devices, which might find 
technological applications in spin-eletronics
and the development of new low-power electronics.
A crucial first step towards the realization of such devices is
the ability to control and manipulate the edge states using, for example,
magnetic fields~\cite{brydonNJP2013,Schnyder13,LinderPRL2010},
electric fields~\cite{EzawaPRL2015,WrayJPCS2013}, or mechanical
strain.

A significant omission in our discussion is the effect of
interactions beyond their role in generating the superconducting
gap. The BdG Hamiltonian, on which the topological
classification is based, is a single-particle theory of the gapped
state. Nevertheless, many candidate topological superconductors are
strongly correlated materials, where interactions may exert a profound
influence over the low temperature physics. The topological 
properties of interacting systems are of much interest, but remain
poorly
understood~\cite{fidkowskiKitaevPRB10,vishwanathPRX13,wangSenthilPRB14,GuWenPRB14}.  A complete classification of  interacting
systems in one dimension has been obtained using matrix product states
\cite{pollmannPRB10,ChenPRB11,schuchPRB11,pollmannPRB12}, but such a
classification is currently missing for higher dimensions. Although
BdG theory generally gives a 
good description of unconventional superconductors (at least at a
phenomenological level), interactions could still qualitatively alter
the edge physics: For example, 
residual interactions may cause an instability 
of the surface states, in particular the flat bands with their
divergent density of states. So far, attention has focused mainly on
instabilities of the surface states of $d$-wave
superconductors~\cite{MatsumotoJPSJ1995,CovingtonPRL1997,HonerkampEPL2000,PotterLeePRL2014},
although the instability of flat bands of spinless superconductors has
also been studied~\cite{LiNJP2013}. 

To conclude, the discovery that electronic systems can have
nontrivial topology has opened a new perspective on the
fundamentals of condensed matter physics. Our review
highlights the large changes wrought by this revolution in the
subfield of nodal superconductivity, producing a new classification
of this disparate class of materials by topological features of their
gap structure. This does not supersede previous classifications, but
rather complements them, and indeed it is already expected that
workers in this field pay attention to topological
properties. We can only see this trend continuing, and that it will
not be long before consideration of topological aspects will 
become ubiquitous and unremarkable in the study of unconventional
superconductivity.

\ack
We would like to thank our colleagues and collaborators
A. Akbari, Y. Asano, C. Ast, H. Benia, P. Biswas, B. Bujnowski,
P.-Y. Chang, W.~Chen, C.-K. Chiu, M. Cuoco, S. Das Sarma, I. Eremin,
M. Fischer, J. Goryo, P. Goswami, I.~Gruzberg, J. Hofmann, P. Horsch,
G. Khaliullin, K.~Kern, L. Klam, D.~Lee, C.-T. Lin, A.~Ludwig,
H.~Luetkens, D. Manske, S.~Matsuura, W. Metzner, C.~Mudry, T. Neupert,
Y. Nohara, A. Ole\'s, J. P. Paglione, D. Peets, R. Queiroz, S. Rex,
B. Roy, S. Ryu, J. D. Sau, M. Sigrist, A. Sudb\o, Z. Sun, R. Thomale,
C.~Timm, P. Wahl, V. M. Yakovenko, and A. Yaresko.  
Special thanks are due to Peter Horsch for his constructive comments
on the manuscript.

\section*{References}

\bibliography{JPCMreviewRefs_v4}

\end{document}